\documentclass[aps,pra,amssymb, amsmath,nobibnotes,nobibnotes,reprint,superscriptaddress,showpacs,showkeys]{revtex4-1}
\usepackage[colorlinks, linkcolor=blue, anchorcolor=blue, citecolor=blue]{hyperref}
\usepackage{graphicx}
\usepackage{revsymb4-1}
\usepackage{bm}
\usepackage{braket}

\begin{document}

\title{Atoms versus photons as carriers of quantum states}

\author{Smail Bougouffa}
\email{sbougouffa@hotmail.com}
\affiliation{Department of Physics, Faculty of Science, Taibah University, P.O. Box 30002, Madinah, Saudi Arabia}
\author{Zbigniew Ficek}
\email{zficek@kacst.edu.sa}
\affiliation{National Center for Mathematics and Physics, KACST, P.O. Box 6086, Riyadh 11442, Saudi Arabia}

\date{\today}

\begin{abstract}
The problem of the complete transfer of quantum states and entanglement in a four-qubit system composed of two single-mode cavities and two two-level atoms is investigated. The transfer of single and double excitation states is discussed for two different coupling configurations between the qubits. In the first, the coupling is mediated by the atoms that simultaneously couple to the cavity modes. In the second configuration, each atom resides inside one of the cavities and the coupling between the cavities is mediated by the overlapping field modes. A proper choice of basis states makes it possible to identify states that could be completely transferred between themselves. 
Simple expressions are derived for the conditions for the complete transfer of quantum states and entanglement. These conditions impose severe constraints on the evolution of the system in the form of constants of motion. The constrains on the evolution of the system imply that not all states can evolve in time, and we find that the evolution of the entire system can be confined into that occurring among two states only. Detailed analysis show that in the case where the interaction is mediated by the atoms, only symmetric superposition states can be completely and reversibly transferred between the atoms and the cavity modes. In the case where the interaction is mediated by the overlapping field modes, both symmetric and antisymmetric superposition states can be completely transferred. 
We also show that the system is capable of generating purely photonic NOON states, but only if the coupling is mediated by the atoms, and demonstrate that the ability to generate the NOON states relies on perfect transfer of an entanglement from the atoms to the cavity~modes.  
\end{abstract}



\maketitle

\section{Introduction}

The ability to control the process of transferring quantum states between remote systems is essential for quantum information processing and the development of quantum computation~\cite{nil,cz97}. The main challenge is the achievement of a quantum interface that would be able to transfer quantum states with hight fidelity. Several schemes have been proposed and experimentally implemented where a high fidelity transfer of quantum state was achieved based on the creation of a strong coupling between the systems~\cite{mo03,ww07,bb07,pa09}. In this limit, the time scale of the transfer process is much shorter than the time scale for dissipation in the system due to a coupling to an external environment. Over this time scale, coherent and reversible transfer of a quantum state can be achieved~\cite{nh12}.

The transfer of a quantum state corresponds to the transfer of correlations from the states of one system to the states of an anther system. In most schemes, linear atomic chains or atomic (spin) lattices are considered for the transfer of quantum states~\cite{kc08,vz12,wm12}. In these models the transfer is mediated by the direct dipole-dipole coupling between neighbouring atoms, which induces the flow of an initial excitation through the chain~\cite{mf05}. 
However, the dipole-dipole coupling may create correlations between the atoms so the state would effectively be transferred through the correlated system. Thus, the final state of the system might not be related in a simple way to the initial state. The atoms can be found in an entangled state even if the initial state was a separable state~\cite{ft02}. 

Other schemes involve a chain of remote cavities each containing a single atom or optical lattice, the so-called Jaynes-Cummings (JC) cells~\cite{JC63,pa63,B96}. The cells could be independent~\cite{IG07,MJ06,MJ07,MJ10,SMZ09}, or coupled to each other through the overlapping field modes~\cite{nr07,oi08,ew10,dz12,gl12,xf12}, connected by a short fibre~\cite{pe97,sm06,pl07,zh09,zy10,sl12}. In these schemes, photons act as carriers for the transfer of quantum states and controlled transfer is implemented by an appropriate choice of the coupling strength of the overlapping field modes. The cavity modes could be in a vacuum state or in a correlated state~\cite{S10,TMSZ10,SA11,S11,SZ11}. 

Alternative to the schemes involving photons are models in which atoms are treated as a small reservoir or a spin bath through which an initial state could be transferred to the field modes~\cite{oo07,lr10,sy11}. The atoms forming the reservoir can be independent or may interact with each other, the two cases that could give different results for the transfer process of a quantum state. 

Using the Schr\"{o}dinger or master equation approach, one can study the transfer process of an initial quantum state for the three classes of systems described above (atomic chains or lattices, coupled JC cells, field modes coupled through atomic reservoir). 
In this paper, we address the question of the complete transfer of quantum states and entanglement in four-qubit systems composed of two single-mode cavities and two two-level atoms. We work in the resonance regime and consider two different coupling configurations between the qubits specified by two distinctly different types of the interaction Hamiltonians. In the first, we assume that both atoms simultaneously couple to two independent cavity modes resulting in an effective system analog of the system composed of two qubits coupled to a small reservoir. Here, the coupling between the field qubits is mediated by the atoms. The other configuration corresponds to a system of two directly coupled JC cells, where the cells are coupled through the overlapping mode functions of the cavity fields. In this case, the coupling between the qubits is mediated by photons. In what follows, we assume that dissipation effects are not present or can be neglected. This restricts our results to the case of a strong coupling between qubits. In fact this is not an overly restrictive limitation regarding the recent progress in the cavity QED technology, where strong couplings have been achieved~\cite{tt98}.

We determine the nature of quantum states that can be {\it completely} transferred under the action of the two different interaction Hamiltonians, and under what conditions the complete transfer of the states corresponds to perfect transfer of entanglement between atoms and the field modes.
Of course, there is a large number of states to which an initial state could be transferred under the action of the interaction Hamiltonians. However, we find that only few of them can be completely transferred between themselves. Moreover, we show that the complete transfer of a quantum state does not necessary mean perfect transfer of entanglement.

The paper is organized as follows. In Sec.~\ref{sec2}, we provide detailed description of the four-qubit system and the coupling configurations between the qubits. A detailed calculation of the probability amplitudes of single excitation states is presented in Sec.~\ref{sec3}, and in Sec.~\ref{sec4} we extend these calculations to double excitation states. We point out the existence of constants of motion that significantly reduce the number of states that could evolve in time. We find that the Hilbert space of the system splits into independent subspaces, each composed of two states only. In other words, the system behaviours in this case as if it were composed of independent two-state subsystems.  
No perfect transfer of an entanglement is achieved even if the states are completely transferred between themselves when one of the states contains correlations between the atoms and the cavity modes. The perfect transfer of the entanglement is achieved between states that are factorable into a product of the atomic and the field states. The effect of losses on the two state evolution and entanglement transfer is briefly discussed in Sec.~\ref{loss}. Finally, we summarize our results in Sec.~\ref{seccon}. Some details of the calculation of the concurrence and the logarithmic negativity of the atomic and the cavity-field subsystems are presented in the Appendix.

\section{General formalism}\label{sec2}

We consider a four qubit system, two identical cavities each composed of a single bosonic mode of frequency~$\omega_{c}$, and two identical atoms modelled as having two energy states, a ground state $\ket{g_{i}}$ and an excited state $\ket{e_{i}}$ separated by frequency $\omega_{0}$ and connected by a transition dipole
moment $\mu_{i}\, (i=A,B)$.  The atoms are represented by spin lowering and raising operators $\sigma_{i}^{-}$
and~$\sigma^{+}_{i}$, whereas the cavity modes are represented by the annihilation and creation operators $(\hat{a},\hat{b})$, and $(\hat{a}^{\dag},\hat{b}^{\dag})$, respectively. We use indexes $(A,B)$ to label the atoms and~$(a,b)$ to label the cavity modes. Each of the qubit pairs, either atoms $(A,B)$ or the cavity modes~$(a,b)$, can be used as a mediator between the qubits.

Our objective is to determine which of the qubits, atoms (fermions) or photons (bosons), are better carriers of quantum states and entanglement from one qubit pair to the other. For this purpose, we concentrate on the transfer of an initial state through the system for two configurations of the coupling between the qubits. In the first configuration, we use atoms as "couplers" by sending them through the cavities. In the second configuration, the atoms are assumed to reside inside the cavities, with one atom in each cavity, and the coupling between the cavities is mediated by overlapping of the evanescent mode functions of the cavity modes. Alternatively, this could be done by connecting the cavities by a short optical fibre. In this kind of coupling, photons act as carriers of an excitation between the cavities.
\begin{figure}[thb]
\begin{center}
\begin{tabular}{c}
\includegraphics[width=0.75\columnwidth]{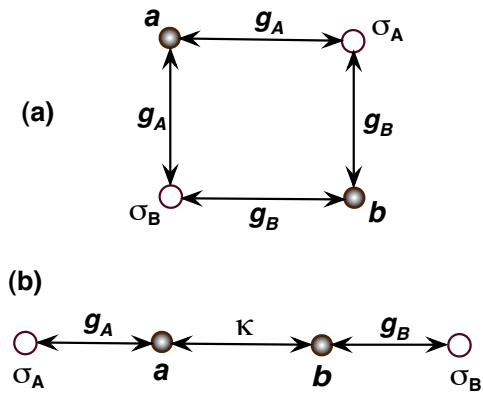}
\end{tabular}
\end{center}
\caption{(Colour online) Schematic diagram of two coupling configurations between qubits in the system composed of two single-mode cavities $(a,b)$ and two two-level atoms $(\sigma_{A}, \sigma_{B})$. In $(a)$, both atoms simultaneously couple to the cavity modes resulting in an effective close loop system. In $(b)$, two JC cells are coupled with a strength $\kappa$ through the overlapping of the mode functions of the cavity fields.} 
\label{sbfig1}
\end{figure}

Figure~\ref{sbfig1} shows two possible coupling configurations in the system. In the first, illustrated in Fig.~\ref{sbfig1}(a), each atom couples to both of the cavity modes that are not directly coupled to each other. In geometrical terms, this configuration is equivalent to a square shape or a closed chain configuration of the four qubits. In the second, illustrated in Fig.~\ref{sbfig1}(b), each atom couples to only one of the modes that are directly coupled to each other. The coupling between the cavity modes could be arranged through the overlap of the mode functions of the cavity fields in the region between the cavities. We see from the figure that the system is equivalent to an open chain of four qubits. This system can be regarded as a double JC systems, two JC systems coupled to each other with the coupling strength~$\kappa$.

The two configurations of the coupling between the cavities can be distinguished by different forms of the interaction Hamiltonian of the atoms and the cavity modes. The total Hamiltonian of the system can be written as
\begin{equation}
\hat{H} = \hat{H}_{0} + \hat{H}_{I} ,\label{e1}
\end{equation}
where
\begin{align}
\hat{H}_{0} = \hbar\omega_{0}\left(\hat{\sigma}_{A}^{z}+\hat{\sigma}_{B}^{z}\right)
+\hbar\omega_{c}(\hat{a}^{\dag}\hat{a}+\hat{b}^{\dag}\hat{b})
\label{e2}
\end{align}
is the free (unperturbed) Hamiltonian of the atoms and the cavity modes, and $\hat{H}_{I}$ is the interaction between the atoms and the cavity modes. The interaction depends on the configuration of the coupling between the atoms and the cavities and between the cavity modes themselves.

In the case of Fig.~\ref{sbfig1}(a), the interaction Hamiltonian, under the rotating-wave approximation, can be written as
\begin{align}
\hat{H}_{I} =&\, \hbar\left[g_{a}\left(\hat{\sigma}_{A}^{+} +\hat{\sigma}_{B}^{+}{\rm e}^{ikR_{AB}}\right)\hat{a}\right. \nonumber\\
&\left. +g_{b}\left(\hat{\sigma}_{A}^{+} +\hat{\sigma}_{B}^{+}{\rm e}^{ikR_{AB}}\right)\hat{b} +{\rm H.c.}\right] .\label{e3}
\end{align}
The Hamiltonian consists of two terms. The first term describes the interaction of the atoms with the field of the cavity $a$, and the second term describe the interaction with the field of the cavity $b$. The strength of the interactions is determined by the coupling coefficients~$g_{a}$ and $g_{b}$, respectively. In practice we may encounter a situation in which atoms traveling through the cavities would be at different positions. Therefore, we have introduced the phase factor $\exp(ikR_{AB})$ in the coupling coefficients reflecting a phase difference in the oscillation of the atomic dipole moments separated by a finite distance~$R_{AB}$. If the atoms are separated by less than a wavelength, $kR_{AB}\ll 1$, the phase difference may be taken to be equal to zero that the atomic dipoles oscillate in phase. When the atoms are separated by a half of the wavelength the phase difference between the atomic dipoles equals $\pi$, the phase factor $\exp(ikR_{AB})=-1$, and then the atomic dipoles oscillate with opposite phases.

In the case of Fig.~\ref{sbfig1}(b), the interaction Hamiltonian may be written as
\begin{align}
\hat{H}_{I} &= \hbar \left(g_{a}\hat{\sigma}_{A}^{+}\hat{a} + g_{b}\hat{\sigma}_{B}^{+}\hat{b} +{\rm H.c.}\right) 
\nonumber\\
&+\hbar \kappa\left(\hat{a}\hat{b}^{\dagger} +\hat{b}\hat{a}^{\dagger}\right) ,\label{e3a}
\end{align}
where $g_{i}\, (i=a,b)$ is the coupling coefficient between the $i$th atom and its local cavity field,  and $\kappa$ is the coupling coefficient between the cavity modes. In the absence of the coupling, $\kappa =0$, the system described by the Hamiltonian~(\ref{e3a}) is completely equivalent to two independent JC models.

The calculations for both configurations are based on the solution of the Schr\"{o}dinger equation for the wave function of the system
\begin{equation}
    i\hbar\frac{\partial}{\partial t}\ket{\Psi_{n}(t)} = \hat{H}\ket{\Psi_{n} (t)} ,\quad n=1,2 ,\label{e4}
\end{equation}
where $n$ labels the number of excitations present in the system. We work in the interaction representation to remove the time dependence at the optical frequencies and expand the wave function in the basis of the eigenstates of the free Hamiltonian~$H_{0}$. 

For the case of a single $(n=1)$ excitation present in the system, the wave function can be expanded in terms of four product state vectors
\begin{equation}
\ket{\Psi_{1}(t)} = \sum_{j=0}^{4}C_{j}(t)\ket{j} ,\label{e5}
\end{equation}
where $\ket{j}$ represents an excitation eigenstate of the free Hamiltonian
\begin{align}
\ket{1} &= \ket{e_{A},g_{B},0_{a},0_{b}} ,\quad \ket{2} =\ket{g_{A},e_{B},0_{a},0_{b}} , \nonumber\\
\ket{3} &= \ket{g_{A},g_{B},1_{a},0_{b}} ,\quad \ket{4} =\ket{g_{A},g_{B},0_{a},1_{b}} ,\label{e4a}
\end{align}
and $C_{j}(t)$ is a slowly varying part of the probability amplitude of the state $j$. The space of the $\ket{j}$ states can be formally divided into two subspaces; one involving states~$\ket 1,\ket 2$ corresponding to one quantum present in either of the atoms, and the other involving states $\ket 3,\ket 4$ corresponding to the atoms residing in their ground states and the excitation present in either of the cavity modes or in a superposition of the modes.  

When two quanta of excitation $(n=2)$ are present in the system, the wave function can be expanded in terms of eight product state vectors
\begin{equation}
\ket{\Psi_{2}(t)} = \sum_{j=0}^{8}D_{j}(t)\ket{\tilde{j}} ,\label{e5a}
\end{equation}
where $\ket{\tilde{j}}$ are two-quanta eigenstates of the free Hamiltonian
\begin{align}
\ket{\tilde{1}} &= \ket{e_{A},e_{B},0_{a},0_{b}} ,\quad \ket{\tilde{2}} =\ket{e_{A},g_{B},0_{a},1_{b}} ,\nonumber\\
\ket{\tilde{3}} &= \ket{g_{A},e_{B},1_{a},0_{b}} ,\quad \ket{\tilde{4}} =\ket{g_{A},g_{B},1_{a},1_{b}} ,\nonumber\\
\ket{\tilde{5}} &= \ket{e_{A},g_{B},1_{a},0_{b}} ,\quad \ket{\tilde{6}} =\ket{g_{A},g_{B},2_{a},0_{b}} ,\nonumber\\
\ket{\tilde{7}} &= \ket{g_{A},e_{B},0_{a},1_{b}} ,\quad \ket{\tilde{8}} =\ket{g_{A},g_{B},0_{a},2_{b}} .\label{e7}
\end{align}
The space of the product states can be formally divided into two subspaces. One involves states $\ket{\tilde{2}},\ket{\tilde{3}},\ket{\tilde{5}},\ket{\tilde{7}}$ corresponding to the two quanta of excitation evenly redistributed over the atomic system and the cavity modes. The other set involves states $\ket{\tilde{1}}, \ket{\tilde{4}}, \ket{\tilde{6}}, \ket{\tilde{8}}$ corresponding to both quanta being either in the atomic system or in the cavity modes.

\section{Transfer of single excitation states}\label{sec3}

Let us first consider the transfer of single excitation states between atoms and the cavity modes. We are particularly interested in which states can be maximally transferred and whether the complete transfer is accompanied by complete transfer of entanglement. The question of the entanglement transfer will be addressed by considering the concurrence of the atomic and the cavity states. This makes it possible to estimate the efficiency of the transfer of the initial state between the atoms and the cavity modes that is mediated by atoms, and compare the transfer efficiency with that mediated by photons.

\subsection{Transfer mediated by atoms}

Consider the configuration in Fig.~\ref{sbfig1}(a), in which the interaction is determined by the Hamiltonian (\ref{e3}). By substituting the wave function (\ref{e5}) into Eq.~(\ref{e4}), we obtain the following system of coupled differential equations for the probability amplitudes
\begin{align}
i\dot{C}_{1} &=  g_{a}C_{3} + g_{b}C_{4} ,\quad 
i\dot{\tilde{C}}_{2} =  g_{a}C_{3} + g_{b}C_{4} ,\nonumber\\
i\dot{C}_{3} &=  -\Delta C_{3} + g_{a}\!\left(C_{1}\!+\!\tilde{C}_{2}\right) ,\nonumber\\
i\dot{C}_{4} &=  -\Delta C_{4} + g_{b}\!\left(C_{1}\!+\!\tilde{C}_{2}\right) ,\label{e8u}
\end{align}
where $\Delta =\omega_{0}-\omega_{c}$ is the detuning of the atomic resonance frequency $\omega_{0}$ from 
the cavity frequency $\omega_{c}$ and $\tilde{C}_{2} =\epsilon^{\ast}C_{2}$, with $\epsilon =\exp(ikR_{AB})$. In order to simplify the notation without loss of generality, we can assume that the coupling strengths $g_{a}$ and $g_b$ are both real.

In order to solve Eqs.~(\ref{e8u}), we find it convenient to introduce linear combinations of the amplitudes,
\begin{align}
W(t) &= \frac{1}{\sqrt{2}}[C_{1}(t) +\tilde{C}_{2}(t)] ,\, U(t) = \frac{1}{\sqrt{2}}[C_{1}(t) -\tilde{C}_{2}(t)] ,\nonumber\\
X(t) &= \frac{g_{a}}{g_{0}}C_{3}(t) +\frac{g_{b}}{g_{0}}C_{4}(t) ,\, Y(t) = \frac{g_{a}}{g_{0}}C_{4}(t) -\frac{g_{b}}{g_{0}}C_{3}(t) ,\label{eq14}
\end{align}
where $g_{0} =\sqrt{g_{a}^{2}+g_{b}^{2}}$. The amplitudes (\ref{eq14}) correspond to orthonormal states
\begin{align}
\ket{w} &= \frac{1}{\sqrt{2}}\left(\ket{e_{A},g_{B}} +\epsilon\ket{g_{A},e_{B}}\right)\otimes\ket{0_{a},0_{b}}  ,\nonumber\\
\ket{u} &= \frac{1}{\sqrt{2}}\left(\ket{e_{A},g_{B}} -\epsilon\ket{g_{A},e_{B}}\right)\otimes\ket{0_{a},0_{b}} ,\nonumber\\
\ket{x} &= \frac{1}{g_{0}}\left(g_{a}\ket{1_{a},0_{b}} + g_{b}\ket{0_{a},1_{b}}\right)\otimes \ket{g_{A},g_{B}} ,\nonumber\\
\ket{y} &= \frac{1}{g_{0}}\left(g_{a}\ket{0_{a},1_{b}} - g_{b}\ket{1_{a},0_{b}}\right)\otimes \ket{g_{A},g_{B}} .\label{eq14a}
\end{align}
These states are product states of purely atomic and cavity fields (photonic) states. In states $\ket w$ and $\ket u$, the atoms are themselves in maximally entangled states, but the field modes are in a factorized state. On the other hand, in states $\ket x$ and $\ket y$, the cavity modes are in non-maximally entangled states and the atoms are in factorized states. This shows that purely atomic entangled states can be transferred or converted into purely photonic $N=1$ NOON states~\cite{nh12,bcs89,bk00,aa10,wm11}. 

In terms of the superposition amplitudes, Eqs.~(\ref{e8u}) transform into a simple set of equations,
\begin{align}
\dot{W}(t) &=  -\sqrt{2}ig_{0}X(t) ,\quad  \dot{X}(t) = i\Delta X(t) -\sqrt{2}ig_{0}W(t) ,\nonumber\\
\dot{Y}(t) &= i\Delta Y(t) ,\quad \dot{U}(t) =  0 ,\label{e8p}
\end{align}
which shows that the amplitudes $X(t)$ and $W(t)$ are coupled only among themselves, whereas $Y(t)$ and $U(t)$ evolve independent of the others.  

The solutions of Eqs.~(\ref{e8p}) are readily found to be 
\begin{align}
W(t) &= W(0)\,{\rm e}^{\frac{1}{2}i\Delta t}\left[\cos(\Omega t) +i\frac{\Delta}{\Omega}\sin(\Omega t)\right] \nonumber\\
&-iX(0)\frac{\sqrt{2}g_{0}}{\Omega}{\rm e}^{\frac{1}{2}i\Delta t}\sin(\Omega t) ,\nonumber\\
X(t) &= X(0)\,{\rm e}^{\frac{1}{2}i\Delta t}\left[\cos(\Omega t) +i\frac{\Delta}{\Omega}\sin(\Omega t)\right] \nonumber\\
&-iW(0)\frac{\sqrt{2}g_{0}}{\Omega}{\rm e}^{\frac{1}{2}i\Delta t}\sin(\Omega t) ,\nonumber \\
Y(t) &= Y(0) {\rm e}^{i\Delta t} ,\quad U(t)  = U(0) ,\label{e9u}
\end{align}
where $\Omega= \sqrt{2g^{2}_{0}+\Delta^{2}/4}$ is a detuned one-photon Rabi frequency. 

The solutions (\ref{e9u}) exhibit a number of interesting features. First, the direct coupling between the amplitudes~$W(t)$ and $X(t)$ indicates that states $\ket w$ and~$\ket x$ form a subspace in the Hilbert space composed of two states that can be reversibly transferred between themselves. Each of the remaining amplitudes $Y(t)$ and $U(t)$ evolves independently of the other states, so that states $\ket y$ and $\ket u$ cannot be accessed from and transferred into other states. Once the system is initially prepared in either $\ket w$ or $\ket x$ state, the state vector of the system will evolve only between these two states. The system will behave in this case as a two-state system. Second, the time evolutions of the atomic and the field states are completely symmetric with $W(t)$ and $X(t)$ interchanged, despite the fact that the atoms are not equally coupled to the cavity modes. Third, the linear superposition~$U(t)$ does not evolve in time. Thus, if initially $U(0)\neq 0$, it will remain unchanged for all time. In other words, the initial population will be trapped in this state and never evolve. Fourth, the time evolution of the superposition amplitudes is independent of $\epsilon$, i.e. it is independent of the relative phase between the atoms.

The presence of the trapping states has a dramatic effect on transfer of entanglement. In order to see it, we evaluate concurrences~\cite{W98} of the atoms and the cavity modes, as described in the Appendix, and find
\begin{align}
 C_{AB}(t) &= \left|\left(|W(t)|^{2} -|U(0)|^{2}\right) +2{\rm Im}[U(0)W^{\ast}(t)]\right| ,\nonumber\\
 C_{ab}(t) &=  2\left|g_{a}g_{b}\left(|X(t)|^{2} -|Y(0)|^{2}\right)\right. \nonumber\\
 &\left. +g_{a}^{2}X(t)Y^{\ast}(t) -g_{b}^{2}X^{\ast}(t)Y(t)\right|/g^{2}_{0} .\label{e9p}
\end{align}
These expressions show explicitly that only initial states with amplitude $U(0)=0$ can be completely transferred between the atoms and the cavity modes. However, the condition of $U(0)=0$ is necessary but not sufficient for the complete transfer of entanglement. It is easy to see, since the factor $2g_{a}g_{b}/g^{2}_{0}$, appearing in the expression for $C_{ab}(t)$ is smaller than one, that it can be easily verified that $C_{ab}(t_{n})$ could be equal to $C_{AB}(0)$ at some particular times $t_{n}$ only if the atoms and the cavity modes have the same frequencies~$(\Delta=0)$ and the atoms couple the the cavities with the same coupling strengths~$(g_{a}=g_{b})$. Needless to say, $U(0)$ must be zero, and the qubits must be indistinguishable for the transfer of the entanglement to be complete. Under these conditions an initial, not necessary maximally entangled state of the atoms, can be completely or perfectly transferred to the cavity modes. 

It is interesting that despite of the complete symmetry between $W(t)$ and $X(t)$, the transfer of maximally entangled states between the atoms and the cavity modes is not symmetric. For example, when the cavity modes were initially in the vacuum state but the atoms were prepared in the entangled state, i.e. the initial state of the system was 
\begin{align}
\ket{\Psi_{0}} = \frac{1}{\sqrt{2}}(\ket{e_{A}g_{B}}+\ket{g_{A}e_{B}})\otimes\ket{0_{a},0_{b}} ,\label{eq11s}
\end{align}
then the concurrence of the cavity modes varies in time~as
\begin{equation}
C_{ab}(t) = \frac{2g_{a}g_{b}}{\left(g_{a}^{2}+g_{b}^{2}\right)}\sin^{2}(\Omega t) .
\end{equation}
On the other hand, when initially atoms were in their ground states but the cavity modes were prepared in the $N=1$ NOON state,
\begin{align}
\ket{\Psi_{0}} = \frac{1}{\sqrt{2}}(\ket{1_{a}0_{b}}+\ket{0_{a}1_{b}})\otimes\ket{g_{A},g_{B}} ,\label{eq12s}
\end{align}
then the concurrence of the atoms varies~as
\begin{equation}
C_{AB}(t) =\frac{(g_{a}+g_{b})^{2}}{2(g_{a}^{2}+g_{b}^{2})}\sin^{2}(\Omega t) .
\end{equation}
Evidently, the variations of the concurrences are not equal, and $C_{AB}(t)\geq C_{ab}(t)$ for all times indicating that the transfer of the initial entanglement from the field modes to the atoms occurs with a better efficiency than the transfer of the entanglement from the atoms to the modes. 

The asymmetry in the entanglement transfer is more dramatic if the cavity modes were initially prepared in a state $\ket x$,  $(X(0)=1)$. In this case, the concurrences are
\begin{align}
C_{AB}(t) = \sin^{2}(\Omega t) ,\quad
C_{ab}(t) = \frac{2g_{a}g_{b}}{\left(g_{a}^{2}+g_{b}^{2}\right)}\cos^{2}(\Omega t) .
\end{align}
We see that the atoms, entering the cavities in their ground states can be maximally entangled independent of whether the cavity modes were initially prepared in the maximally entangled state or not.  

The reason for this feature lies in the fact that the system of two atoms unequally coupled to two cavities effectively behaves as a system composed of two atoms equally coupled to a single superposition mode. 
To show this, we introduce the collective symmetric atomic operator
\begin{equation}
\hat{\sigma}_{s}^{+} =  \left(\sigma_{A}^{+} +\epsilon \sigma_{B}^{+}\right) ,
\end{equation}
together with symmetric and antisymmetric superposition operators of the field modes 
\begin{equation}
\hat{d}_{1} = \left(g_{a}\hat{a} + g_{b}\hat{b}\right)\!/g_{0} ,\quad \hat{d}_{2} = \left(g_{a}\hat{b} - g_{b}\hat{a}\right)\!/g_{0} ,
\end{equation}
and find that in terms of the superposition operators the Hamiltonian (\ref{e3}) takes a simple form
\begin{align}
H_{I} = \hbar g_{0}\hat{\sigma}_{s}^{+}\hat{d}_{1} +{\rm H.c.} \label{e3x}
\end{align} 
We see that the atoms effectively interact only with the symmetric mode $\hat{d}_{1}$. The antisymmetric mode $\hat{d}_{2}$ is completely decoupled from the atoms. Thus, in terms of the superposition operators, the system effectively behaves as a single JC system composed of two atoms equally coupled to a single mode $\hat{d}_{1}$. In terms of the state transfer, it means that only symmetric states can be transferred between the atoms and the cavity modes.

\subsection{Transfer mediated by photons}

We now consider the configuration in Fig.~\ref{sbfig1}(b) in which each atom is located at a fixed position inside one of the cavities and the coupling between the cavities is mediated by overlapping cavity modes. In this case, an excitation is carried by photons. In practice this system can be realized by using two optical cavities each containing a single atom and the coupling between the cavities mediated by an optical fibre~\cite{pe97,sm06,pl07,zh09,zy10,sl12}. Alternatively, one can use two optical traps each containing a single atom and the coupling maintained by the photon tunneling effect~\cite{rw06,gr11,uc11} or, in the case of circuit QED, capacitive or inductive coupling~\cite{bm03}.

Before considering the process of transferring quantum states between the qubits, let us first analyse the Hamiltonian of the system in which the interaction is described by Eq.~(\ref{e3a}). Introducing symmetric and antisymmetric linear combinations of the atomic and field operators
\begin{align}
\hat{\sigma}^{+}_{s} &= \left(g_{a}\hat{\sigma}^{+}_{A}+g_{b}\hat{\sigma}^{+}_{B}\right)\!/g_{0} ,\quad
\hat{\sigma}^{+}_{a} = \left(g_{a}\hat{\sigma}^{+}_{A} -g_{b}\hat{\sigma}^{+}_{B}\right)\!/g_{0} ,\nonumber\\
\hat{d}^{\dag} &= \left(\hat{a}^{\dag} +\hat{b}^{\dag}\right)\!/\sqrt{2} ,\quad \hat{c}^{\dag} = \left(\hat{a}^{\dag} -\hat{b}^{\dag}\right)\!/\sqrt{2} ,
\end{align}
we find that the total Hamiltonian of the system can be written~as
\begin{align}
\hat{H} &= \hbar\omega_{0}\left(\sigma_{A}^{z}+\sigma_{B}^{z}\right)
+\hbar(\omega_{c} +\kappa)\hat{d}^{\dag}\hat{d} +\hbar(\omega_{c} -\kappa)\hat{c}^{\dag}\hat{c}\nonumber\\
&+\frac{\hbar g_{0}}{\sqrt{2}}\left(\hat{d}^{\dag}\hat{\sigma}^{-}_{s} +\hat{c}^{\dag}\hat{\sigma}^{-}_{a} + \text{H.c.}\right) .\label{hc}
\end{align}
The first line of the Hamiltonian (\ref{hc}) represents the free energy of the atoms and the superposition modes with the energies of the superposition modes separated by $2\kappa$; the coupling of the field modes thus lifts their degeneracy. The second line of Eq.~(\ref{hc}) represents the interaction of the superposition field modes with the collective atomic systems.

We note that the structure of the Hamiltonian (\ref{hc}) differs significantly from the Hamiltonian we encountered in Eq.~(\ref{e3x}), the counterpart for the coupling mediated by the atoms. The most interesting difference is that two JC systems coupled to each other by overlapping cavity modes can be viewed as two independent and non-degenerate JC systems, one composed of the symmetric modes and the other composed of the antisymmetric modes. This suggests that in this case both symmetric and antisymmetric states could be maximally transferred between the atoms and the cavity modes. The transfers could occur at two different frequencies, the symmetric states could be maximally transferred at frequency $\omega_{a}=\omega_{c}+\kappa$, whereas the antisymmetric states could be transferred at $\omega_{a}=\omega_{c}-\kappa$.

Since $\hat{d}^{\dag}$ and $\hat{c}^{\dag}$ are in the form of the equally weighted linear combinations of the field operators whereas $\hat{\sigma}^{+}_{s}$ and $\hat{\sigma}^{+}_{a}$ are in general not equally weighted combinations of the atomic operators, one can then show that maximally entangled states of the field modes can be created in the system even if the atoms are weakly entangled or even separable. This feature is opposite of what we encountered in the case of the transfer mediated by the atoms, where maximally entangled states between the atoms were created with separable cavity modes. 

We now turn to the evaluation of the state vector of the system. With the interaction Hamiltonian~(\ref{hc}) the Schr\"odinger equation leads to the following equations of motion for the probability amplitudes
\begin{align}
i\dot{C}_{1} &=  g_{a}C_{3} ,\quad  i\dot{C}_{2} =  g_{b}C_{4} ,\nonumber\\
i\dot{C}_{3} &= -\Delta C_{3} +\kappa C_{4} +g_{a}C_{1} ,\nonumber\\
i\dot{C}_{4} &=  -\Delta C_{4} +\kappa C_{3} +g_{b}C_{2} .\label{e26u}
\end{align}
Equations (\ref{e26u}) can be easily solved and interpreted by rewriting them as equations of motion for symmetric and antisymmetric superpositions 
\begin{align}
C_{s} &= (C_{3} + C_{4})/\sqrt{2} ,\quad C_{a} = (C_{3} - C_{4})/\sqrt{2} ,\nonumber\\
C_{+} &= (g_{a}C_{1} + g_{b}C_{2})/g_{0} ,\quad C_{-} = (g_{b}C_{1} - g_{a}C_{2})/g_{0} .
\end{align}
They are
\begin{align}
i\dot{C}_{s} &= -(\Delta -\kappa)C_{s} +\frac{g_{0}}{\sqrt{2}}C_{+} ,\nonumber\\
i\dot{C}_{a} &=  -(\Delta +\kappa)C_{a} +\frac{w}{\sqrt{2}}C_{-} + \frac{u}{\sqrt{2}}C_{+} ,\nonumber\\
i\dot{C}_{+} &=  \frac{g_{0}}{\sqrt{2}}C_{s} + \frac{u}{\sqrt{2}}C_{a} ,\nonumber\\
i\dot{C}_{-} &=  \frac{w}{\sqrt{2}}C_{a} ,\label{e26m}
\end{align}
where $u = (g_{a}^{2} - g_{b}^{2})/g_{0}$ and $w=2g_{a}g_{b}/g_{0}$. It is clear that the symmetric and antisymmetric modes oscillate at different frequencies. Moreover, in the case of identical JC systems $(u=0)$, the modes evolve independently from each other. However, the modes may evolve independently even if $u\neq 0$. It happens when the field modes are well separated in frequency, i.e. when $\kappa \gg g_{0}$. In this case, the coupling which exists in general between the symmetric and antisymmetric modes is effectively quite weak and can be ignored. It is easy to see. By choosing a new rotating frame with
\begin{align}
\tilde{C}_{s} &= C_{s}{\rm e}^{-i(\Delta -\kappa)t} ,\quad \tilde{C}_{a} = C_{a}{\rm e}^{-i(\Delta +\kappa)t} ,\nonumber\\
\tilde{C}_{+} &= C_{+}{\rm e}^{-i(\Delta -\kappa)t} ,\quad \tilde{C}_{-} = C_{-}{\rm e}^{-i(\Delta +\kappa)t} ,
\end{align}
the equations (\ref{e26m}) can than be written in the form
\begin{align}
i\dot{\tilde{C}}_{s} &= \frac{g_{0}}{\sqrt{2}}\tilde{C}_{+} ,\nonumber\\
i\dot{\tilde{C}}_{a} &=  \frac{w}{\sqrt{2}}\tilde{C}_{-} + \frac{u}{\sqrt{2}}\tilde{C}_{+}{\rm e}^{-2i\kappa t},\nonumber\\
i\dot{\tilde{C}}_{+} &= (\Delta -\kappa)\tilde{C}_{+} + \frac{g_{0}}{\sqrt{2}}\tilde{C}_{s} + \frac{u}{\sqrt{2}}\tilde{C}_{a}{\rm e}^{2i\kappa t} ,\nonumber\\
i\dot{\tilde{C}}_{-} &= (\Delta + \kappa)\tilde{C}_{-} + \frac{w}{\sqrt{2}}\tilde{C}_{a}  .\label{e26n}
\end{align}
Clearly, all the $u$ dependent terms are accompanied by the exponential factors $\exp(\pm 2i\kappa t)$ which in the limit of~$\kappa\gg g_{0}$ rapidly oscillate in time and thus can be ignored. This has the effect of decoupling the equations for the pair of amplitudes $(\tilde{C}_{s},\tilde{C}_{+})$ from the pair $(\tilde{C}_{a},\tilde{C}_{-})$.

The assumption of $\kappa\gg g_{0}$ appears to be practical. For example, in the experiments involving a short fibre coupling the cavities~\cite{sm06,th05}, the coupling strengths of $\kappa\approx 100g_{0}$ can be achieved and $\kappa$ can be increased by decreasing the reflectivity of the cavity mirror connected to the fibre.  

Assuming that $\kappa\gg g_{0}$, i.e. that the modes oscillate at significantly different frequencies, we then readily find analytical expressions for the concurrences, which are of the form
\begin{align}
C_{AB}(t) &= \frac{2g_{a}g_{b}}{g^{2}_{0}}\left||C_{+}(t)|^{2} - |C_{-}(t)|^{2}\right| ,\nonumber\\
C_{ab}(t) &= \left||C_{s}(t)|^{2} - |C_{a}(t)|^{2}\right| ,
\end{align}
where
\begin{align}	
C_{s}(t) &= {\rm e}^{\frac{1}{2}i(\Delta -\kappa)t}\left[C_{s}(0)\cos\frac{1}{2}\Omega_{\kappa}t\right. \nonumber\\
&\left. +i\frac{(\Delta -\kappa)C_{s}(0) -\sqrt{2}g_{0}C_{+}(0)}{\Omega_{\kappa}}\sin\frac{1}{2}\Omega_{\kappa}t \right] ,\nonumber\\
C_{+}(t) &= {\rm e}^{\frac{1}{2}i(\Delta -\kappa)t}\left[C_{+}(0)\cos\frac{1}{2}\Omega_{\kappa}t\right. \nonumber\\
&\left. -i\frac{(\Delta -\kappa)C_{+}(0) +\sqrt{2}g_{0}C_{s}(0)}{\Omega_{\kappa}}\sin\frac{1}{2}\Omega_{\kappa}t \right] ,
\end{align}
with $\Omega_{\kappa} =\sqrt{2g^{2}_{0} +(\Delta -\kappa)^{2}}$, and the amplitudes $C_{a}(t)$ and $C_{-}(t)$ are obtained from the above by changing $s\rightarrow a$, $+\rightarrow -$, $\kappa\rightarrow -\kappa$, and $g\rightarrow w$.  Note that the time evolutions of the field states, determined by $C_{s}(t)$, and the atomic states, determined by $C_{+}(t)$, are not completely symmetric. The evolutions are completely symmetric only at $\Delta=\kappa$. 

Having the explicit forms of the concurrences we now proceed to consider the process of transferring the initial maximally entangled states between the atoms and the cavity modes. Provided the atoms are prepared in the maximally entangled state~(\ref{eq11s}), the initial conditions are~$C_{s}(0)=C_{a}(0)=0$, $C_{+}(0) =(g_{a}+g_{b})/(\sqrt{2}g_{0}) $, and $C_{-}(0) =(g_{a}-g_{b})/(\sqrt{2}g_{0})$. Under this condition, the concurrence~$C_{ab}(t)$ takes the form
\begin{align}
C_{ab}(t) &= \left|\frac{(g_{a}+g_{b})^{2}}{2g^{2}_{0} +(\Delta -\kappa)^{2}}\sin^{2}\left(\frac{1}{2}\sqrt{2g^{2}_{0} +(\Delta -\kappa)^{2}}t\right)\right. \nonumber\\
&\left. -\frac{(g_{a}-g_{b})^{2}}{2g^{2}_{0} +(\Delta +\kappa)^{2}}\sin^{2}\left(\frac{1}{2}\sqrt{2g^{2}_{0} +(\Delta +\kappa)^{2}}t\right)\right| .\label{eq30f}
\end{align}
It is clear by inspection of Eq.~(\ref{eq30f}) that in general, the concurrence is smaller than one for all times $t$. However, the concurrence could attain the maximal value when $\Delta=\kappa$ and $g_{a}=g_{b}=g$, in which case $C_{ab}(t_{n})=1$ at the particular times $gt_{n}=n\pi/2,\, n=1,3,\ldots$. Thus, the cells must be identical and $\Delta=\kappa$ for the transfer of the maximally entangled state from the atoms to the cavity modes to be complete. 

If initially the cavity modes are prepared in the maximally entangled $N=1$ "NOON" state (\ref{eq12s}), which implies $C_{+}(0)=C_{-}(0)=C_{a}(0)=0$ and  $C_{s}(0) =1$, the concurrence~$C_{AB}(t)$ is
\begin{align}
C_{AB}(t) = \frac{4g_{a}g_{b}}{2g^{2}_{0} +(\Delta\!-\!\kappa)^{2}}\sin^{2}\!\left(\frac{1}{2}\sqrt{2g^{2}_{0} +(\Delta -\kappa)^{2}}t\right) .
\end{align}
We see that similar to the case of transferring the maximally entangled state from the atoms to the cavity modes, the transfer of the maximally entangled state from the field modes to the atoms can be complete only if $g_{a}=g_{b}$ and $\Delta =\kappa$.
\begin{figure}[ht]
\includegraphics[width=0.49\columnwidth]{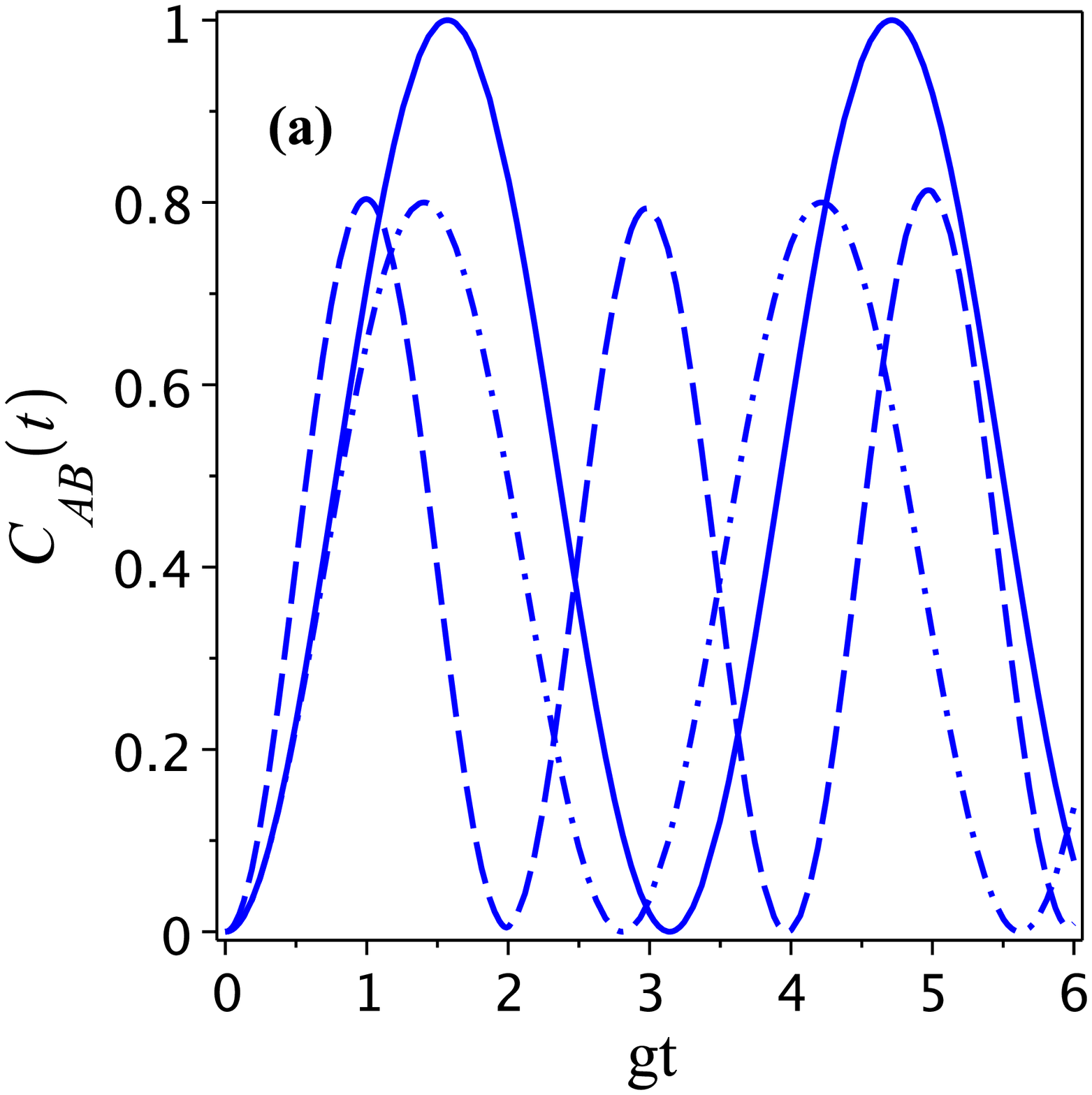}
\includegraphics[width=0.49\columnwidth]{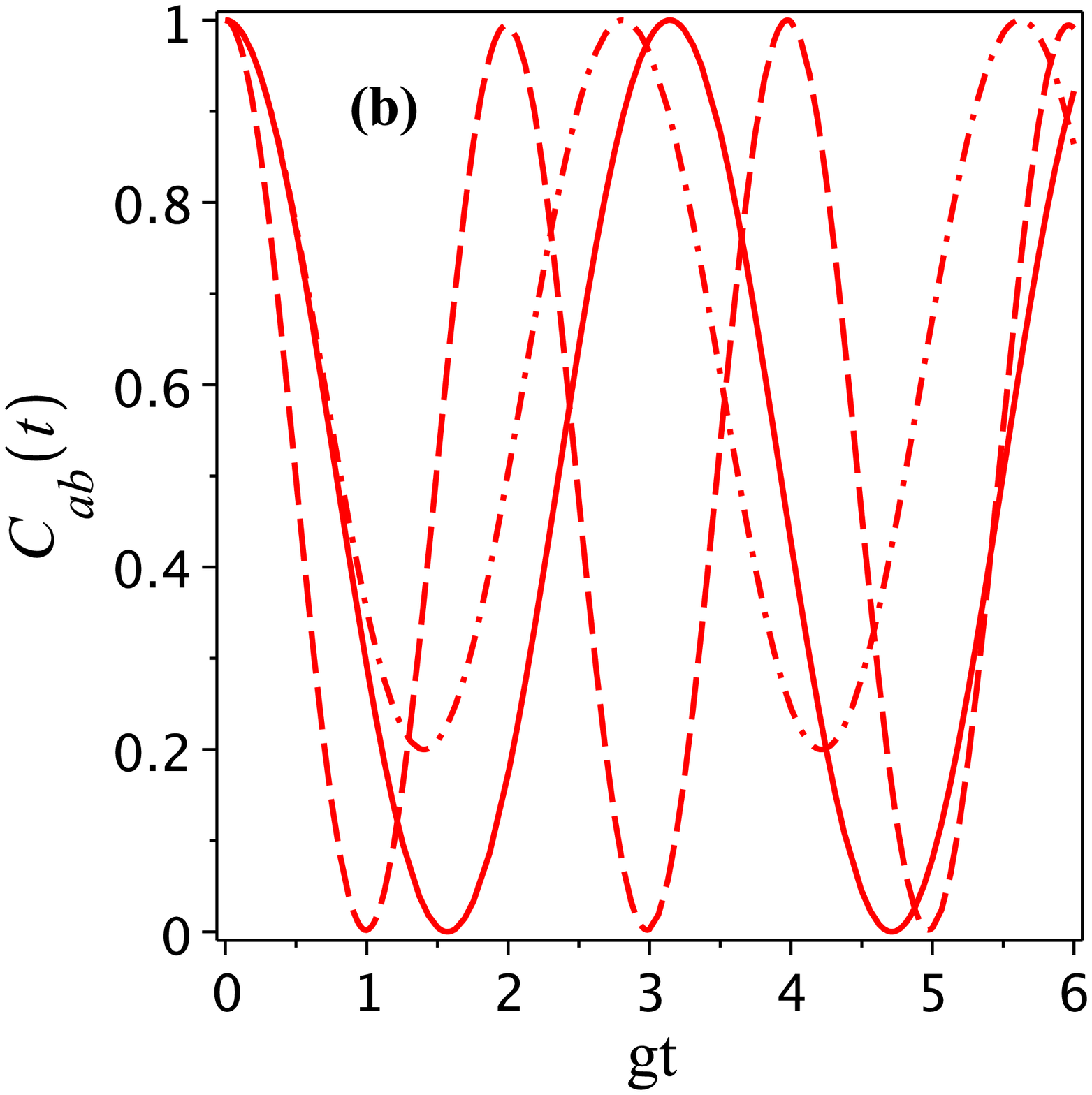}
\caption{(Color online) The time evolution of the concurrences (a) $C_{AB}(t)$ and (b) $C_{ab}(t)$ for $g_{a}=g_{b}=g, \Delta=\kappa$ (solid line), $g_{a}=g, g_{b}=2g, \Delta=\kappa$ (dashed line), and $g_{a}=g_{b}=g, \Delta=5g, \kappa=4g$ (dashed-dotted line). The system was initially in the state described by Eq.~(\ref{eq12s}).}
\label{sbfig2}
\end{figure}

The above considerations are illustrated in Fig.~\ref{sbfig2}, which shows the concurrences $C_{AB}(t)$ and $C_{ab}(t)$ as a function of time for several different values of the parameters $g_{a}, g_{b}, \Delta$ and $\kappa$. It is seen that at certain times and only for $g_{a}=g_{b}$ and $\Delta=\kappa$ the entanglement is completely transferred from the field modes to the atoms. The transfer is far from complete when $g_{a}\neq g_{b}$ and/or $\kappa$ close but not equal to $\Delta$. Notice that independent of the relation between $g_{a}$ and $g_{b}$ and whether $\Delta=\kappa$ or not, the cavity modes become maximally entangled whenever the atoms are disentangled. 

In summary of this section, comparing the results of transferring quantum states and entanglement in the configuration in Fig.~\ref{sbfig1}(b) to those of the configuration in Fig.~\ref{sbfig1}(a), we find that their operations share many common features. For example, in both configurations the coupling strengths of the atoms to the cavity modes must be equal for the transfer of maximally entangled states to be complete. However, there are some important differences. In the configuration in Fig.~\ref{sbfig1}(a), only the symmetric states could be completely transferred, whereas in the configuration in Fig.~\ref{sbfig1}(b) the transfer of both symmetric and antisymmetric states is possible. Moreover, in the configuration in Fig.~\ref{sbfig1}(a) the complete transfer occurs at the exact resonance between the atomic and the cavity frequencies, $\Delta=0$, whereas in the configuration in Fig.~\ref{sbfig1}(b) the optimal transfers occur at nonzero detunings~$\Delta=\pm\kappa$.

\section{Transfer of double excitation states}\label{sec4}

Let us now examine the problem of transferring double excitation states with two quanta present in the system. The two quanta could be in the atoms or in the field modes, or could be shared between the atoms and the field modes. As one can notice from the form of the state vector of the system, Eq.~(\ref{e7}), the presence of two quanta of excitation increases the number of states in which the system can be initially prepared and to which it could evolve. Following the same procedure as in the previous section, we consider separately the transfer of initial states mediated by the atoms and by photons.

\subsection{Transfer mediated by atoms}

In order to discuss the transfer of double excitation states, we make use of the wave function (\ref{e5a}). Since for the transfer mediated by the atoms, the best result is expected to occur when the atomic transition frequencies are in resonance with the cavity field frequencies, we set $\Delta=0$ in what follows in this section. When the wave function (\ref{e5a}) is inserted into Eq.~(\ref{e4}) and we use the interaction Hamiltonian (\ref{e3}), we obtain the following system of coupled differential equations for the probability amplitudes
\begin{align}
i\dot{\tilde{D}}_{1} &= g_{a}(\tilde{D}_{3}+D_{5})+g_{b}(D_{2}+\tilde{D}_{7}) ,\nonumber\\
i\dot{D}_{2} &= g_{b}\tilde{D}_{1}+g_{a}D_{4}+\sqrt{2}g_{b}D_{8} ,\nonumber\\
i\dot{\tilde{D}}_{3} &= g_{a}\tilde{D}_{1} + g_{b}D_{4}+\sqrt{2}g_{a}D_{6} ,\nonumber\\
i\dot{D}_{4} &= g_{a}(D_{2}+\tilde{D}_{7})+g_{b}(\tilde{D}_{3} + D_{5}) ,\nonumber\\
i\dot{D}_{5} &= g_{a}\tilde{D}_{1}+g_{b}D_{4}+\sqrt{2}g_{a}D_{6} ,\nonumber\\
i\dot{D}_{6} &= \sqrt{2}g_{a} (\tilde{D}_{3} + D_{5}) ,\nonumber\\
i\dot{\tilde{D}}_{7} &= g_{b}\tilde{D}_{1} + g_{a}D_{4}+\sqrt{2}g_{b}D_{8} ,\nonumber\\
i\dot{D}_{8} &= \sqrt{2}g_{b}(D_{2}+\tilde{D}_{7}) ,\label{e8}
\end{align}
where $\tilde{D}_{i}=\epsilon^{\ast}D_{i},\, (i=1,3,7)$.

\subsubsection{Independent Jaynes-Cummings systems}

Before continuing with the solution of Eq.~(\ref{e8}), we briefly comment about the difference between the system considered here and that of two independent JC systems, extensively studied by other authors~\cite{MJ06,MJ07,MJ10,SMZ09}.  When each cavity contains one atom, and the cavities are not coupled to each other, we then have two independent JC systems. In this case, the system of equations for the probability amplitudes, Eq.~(\ref{e8}), breaks up into three uncoupled sets of equations. One composed of four coupled equations
\begin{align}
i\dot{\tilde{D}}_{1} &= g_{a}\tilde{D}_{3}+g_{b}D_{2} ,\quad 
i\dot{D}_{2} = g_{a}D_{4}+g_{b}\tilde{D}_{1} ,\nonumber\\
i\dot{\tilde{D}}_{3} &= g_{a}\tilde{D}_{1} +g_{b}D_{4} ,\quad
i\dot{D}_{4} = g_{a}D_{2} +g_{b}\tilde{D}_{3} ,\label{e8a}
\end{align}
describing the evolution of the states with a single excitation present in each of the JC systems, and two separate sets each composed of two coupled equations
\begin{align}
i\dot{D}_{5} = \sqrt{2}g_{a}D_{6} ,\quad 
i\dot{D}_{6} = \sqrt{2}g_{a}D_{5} ,\label{e8b}
\end{align}
and
\begin{align}
i\dot{D}_{7} =  \sqrt{2}g_{b}D_{8} ,\quad 
i\dot{D}_{8} = \sqrt{2}g_{b}D_{7} ,\label{e8c}
\end{align}
describing the evolution of the states with two excitations present in either of the JC systems.

The set of four equations (\ref{e8a}) can be further split into two independent sets by introducing symmetric and antisymmetric superpositions 
\begin{align} 
D_{s1} &= (\tilde{D}_{1} + D_{4})/\sqrt{2} ,\quad D_{s2} = (D_{2} + \tilde{D}_{3})/\sqrt{2} ,\nonumber\\
D_{a1} &= (\tilde{D}_{1} - D_{4})/\sqrt{2} ,\quad D_{a2} = (D_{2} - \tilde{D}_{3})/\sqrt{2} ,\label{e9k}
\end{align}
for which Eqs.~(\ref{e8a}) transform into two separate sets each composed of two coupled equations
\begin{align}
i\dot{D}_{s1} = (g_{a}+g_{b})D_{s2} ,\quad i\dot{D}_{s2} = (g_{a}+g_{b})D_{s1} ,
\end{align}
and 
\begin{align}
i\dot{D}_{a1} = (g_{a}-g_{b})D_{a2} ,\quad i\dot{D}_{a2} = (g_{a}-g_{b})D_{a1} .
\end{align}
The amplitudes (\ref{e9k}) correspond to a set of orthonormal states
\begin{align}
\ket{s1} &= \frac{1}{\sqrt{2}}\left(\ket{e_{A},e_{B},0_{a},0_{b}} +\ket{g_{A},g_{B},1_{a},1_{b}}\right)  ,\nonumber\\
\ket{s2} &= \frac{1}{\sqrt{2}}\left(\ket{e_{A},g_{B},0_{a},1_{b}} + \ket{g_{A},e_{B},1_{a},0_{b}}\right) ,\nonumber\\
\ket{a1} &= \frac{1}{\sqrt{2}}\left(\ket{e_{A},e_{B},0_{a},0_{b}} -\ket{g_{A},g_{B},1_{a},1_{b}}\right) ,\nonumber\\
\ket{a2} &= \frac{1}{\sqrt{2}}\left(\ket{e_{A},g_{B},0_{a},1_{b}} - \ket{g_{A},e_{B},1_{a},0_{b}}\right) .\label{e9m}
\end{align}

We see that despite of the complexity, the dynamics of the system simplifies to that occurring among states of four independent doublets
\begin{align}
\ket{s1}\leftrightarrow \ket{s2},\,  \ket{a1}\leftrightarrow \ket{a2},\, \ket{5}\leftrightarrow \ket{6},\, \ket{7}\leftrightarrow \ket{8} .
\end{align}
This shows explicitly that the Hilbert space of the system splits into four independent subspaces each composed of two states only. In other words, the system behaves in this case as if composed of four independent two-state subsystems. The states of different subsystems oscillate at different Rabi frequencies. For example, the states of the doublet $(\ket{s1}, \ket{s2})$ oscillate at the Rabi frequency $g_{a}+g_{b}$. The states of the remaining subsystems oscillate at frequencies $g_{a}-g_{b}$, $\sqrt{2}g_{a}$, and $\sqrt{2}g_{b}$, respectively. Note that in the limit of~$g_{a}=g_{b}$, the states of the doublet $(\ket{a1}, \ket{a2})$ decouple from each other and become trapping states.

The states (\ref{e9m}) are entangled states of the atoms with the field modes. However, in these states both, the atoms and the field modes are themselves disentangled. It is easy to see. 
Properties of the atoms alone can be described by a reduced density operator
obtained by tracing the total density operator $\rho(t)$ over the cavity modes
\begin{align}
\rho_{AB}(t) &= {\rm Tr}_{ab}\rho(t) = \bra {0_{a}0_{b}} \rho(t) \ket {0_{a}0_{b}} \nonumber\\
&+\bra{1_{a}0_{b}}\rho(t)\ket{1_{a}0_{b}} +\bra {0_{a}1_{b}}\rho(t) \ket{0_{a}1_{b}} \nonumber\\
& +\bra{1_{a}1_{b}}\rho(t)\ket{1_{a}1_{b}} +\bra {2_{a}0_{b}} \rho(t) \ket {2_{a} 0_{b}} \nonumber\\
&+\bra {0_{a} 2_{b}} \rho(t) \ket {0_{a} 2_{b}} .
\end{align}
It is easy to show that in the basis spanned by four state vectors, $\ket{e_{A}e_{B}},\ket{e_{A}g_{B}},\ket {g_{A}e_{B}}, \ket{g_{A}g_{B}}$, the density operator~$\rho_{AB}(t)$ is diagonal with all off-diagonal elements (coherences) equal to zero. Consequently, there is no possibility for entanglement between the atoms.

Yonac {\it et al.}~\cite{MJ06,MJ07} have proposed to include an auxiliary zero-excitation state, $\ket 0 = \ket{g_{A},g_{B},0_{a},0_{b}}$ that permits the two-photon coherence to be involved in the atomic dynamics. When the state $\ket 0$ is included, the following density matrix~$\rho_{AB}(t)$ is obtained
\begin{align}
    \rho_{AB}(t)=\left(
\begin{array}{cccc}
  \rho_{11}(t) & 0 & 0 & \rho_{14}(t) \\
  0 & \rho_{22}(t) & 0 & 0 \\
  0 & 0 & \rho_{33}(t) & 0 \\
  \rho_{41}(t) & 0 & 0 & \rho_{44}(t) \\
\end{array}
\right) ,\label{11ua}
\end{align}
where $\rho_{11}(t)=|D_{1}(t)|^{2}, \rho_{22}(t) =|D_{2}(t)|^{2}+|D_{5}(t)|^{2}$, $\rho_{33}(t) = |D_{3}(t)|^{2}+|D_{7}(t)|^{2}$, $\rho_{44}(t)= |D_{4}(t)|^{2}+|D_{6}(t)|^{2}+|D_{8}(t)|^{2}+|D_{0}(t)|^{2}$, $\rho_{14}(t) = D_{1}(t)D_{0}(t)$, and $D_{0}(t)$ is the amplitude of the auxiliary state $\ket 0$. 

The coherence may result in an entanglement between the atoms. It is easy to see, the concurrence evaluated from the matrix~(\ref{11ua}) takes the form
\begin{equation}
    C_{AB}(t) = {\rm max}\{0,{\cal C}_{2}(t)\} ,
\end{equation}
where
\begin{align}
    {\cal C}_{2}(t) &= 2|\rho_{14}(t)| -2\left(|D_{2}(t)|^{2}+|D_{3}(t)|^2\right) .
\end{align}
It is seen that ${\cal C}_{2}(t)$ can be positive. Thus, the inclusion of~$\ket 0$ into the atomic dynamics is critical for entanglement between two atoms placed inside independent cavities.

We may summarize that the Hilbert space of the independent JC systems can be spanned in terms of four independent subspaces, each composed of two states that can be exchanged between themselves during the evolution. In the case of equal coupling strengths of the atoms to the field modes, one of the subspaces is composed of trapping states that do not evolve.

\subsubsection{Coupled Jaynes-Cummings systems}

We now proceed to solve the set of Eqs.~(\ref{e8}), which shall allow us to discuss the transfer efficiency of doubly excited states in the system composed of two coupled~JC systems. Since the transfer process can be strongly affected by the presence of trapping states in the system, we solve Eqs.~(\ref{e8}) by finding a constant of motion that determine states which do not evolve in time. This allows us to predict which initial states can be completely transferred in the system.

We first observe that $ \dot{\tilde{D}}_{7} = \dot{D}_{2}$ and $\dot{\tilde{D}}_{3} = \dot{D}_{5}$. This suggests the introduction of symmetric and antisymmetric superpositions of the states with the excitation shared between the atoms and the cavity modes
\begin{align} 
D_{s1} &= (\tilde{D}_{3} + D_{5})/\sqrt{2} ,\quad D_{s2} = (D_{2} + \tilde{D}_{7})/\sqrt{2} ,\nonumber\\
D_{a1} &= (\tilde{D}_{3} - D_{5})/\sqrt{2} ,\quad D_{a2} = (D_{2} - \tilde{D}_{7})/\sqrt{2} ,\label{e9}
\end{align}
and find that then the set of Eqs.~(\ref{e8}) splits into two separate sets, one involving six coupled differential equations 
\begin{align}
i\dot{\tilde{D}}_{1} &= \sqrt{2}(g_{a}D_{s1} + g_{b}D_{s2}) ,\nonumber\\
i\dot{D}_{s1} &= \sqrt{2}(g_{b}\tilde{D}_{1} + g_{a}D_{4}+\sqrt{2}g_{b}D_{6}) ,\nonumber\\
i\dot{D}_{s2} &= \sqrt{2}(g_{a}\tilde{D}_{1} + g_{b}D_{4}+\sqrt{2}g_{a}D_{8}) ,\nonumber\\
i\dot{D}_{4} &= \sqrt{2}(g_{a}D_{s2} +g_{b}D_{s1}) ,\nonumber\\
i\dot{D}_{6} &= 2g_{a}D_{s1} ,\nonumber\\
i\dot{D}_{8} &= 2g_{b}D_{s2} ,\label{e10}
\end{align}
and the other composed of two constants of motion
\begin{align} 
i\dot{D}_{a1} = 0 ,\quad i\dot{D}_{a2} = 0 .\label{e11}
\end{align}

Equations~(\ref{e10}) are cumbersome due to the asymmetry in general between the coupling constants of the atoms to the field modes, $g_{a}\neq g_{b}$. However, we have seen in the case of single-quantum states that  the complete transfer of the states and the optimal transfer of entanglement occur when the atoms couple to both cavities with the same strengths, $g_{a}=g_{b}$. Therefore, in what follows we restrict the calculations to the case of~$g_{a}=g_{b}\equiv g$.

When $g_{a}=g_{b}$, additional constants of motion appear. To see this, we introduce symmetric and antisymmetric superposition of the probability amplitudes
\begin{align} 
D_{u} &= (D_{s1} + D_{s2})/\sqrt{2} ,\quad D_{w} = (D_{s1} - D_{s2})/\sqrt{2} ,\nonumber\\
D_{p} &= (D_{6} + D_{8})/\sqrt{2} ,\quad D_{q} = (D_{6} - D_{8})/\sqrt{2} .\label{e12}
\end{align}
We find that the set of coupled equations~(\ref{e10}) can be split into two separate sets, one involving four coupled equations
\begin{align}
i\dot{\tilde{D}}_{1} &= 2gD_{u} ,\nonumber\\
i\dot{D}_{u} &= 2g\tilde{D}_{1} + 2gD_{4}+2gD_{p} ,\nonumber\\
i\dot{D}_{4} &= 2gD_{u} ,\nonumber\\
i\dot{D}_{p} &= 2gD_{u} ,\label{e13}
\end{align}
and the other involving two coupled equations
\begin{align}
i\dot{D}_{w} = 2gD_{q} ,\quad 
i\dot{D}_{q} = 2gD_{w} .\label{e14}
\end{align}
The set of equations (\ref{e13}) can be reduced further to a set of two coupled equations
\begin{align}
i\dot{D}_{u} = 2\sqrt{3}gD_{z} ,\quad
i\dot{D}_{z} = 2\sqrt{3}gD_{u} ,\label{ee15}
\end{align}
and two constants of motion
\begin{align} 
i\dot{D}_{m} = 0 ,\quad i\dot{D}_{n} = 0 ,\label{e16}
\end{align}
where
\begin{align} 
D_{m} = (2D_{p} - D_{1}-D_{4})/\sqrt{6} ,\quad D_{n} = (\tilde{D}_{1} - D_{4})/\sqrt{2} ,\label{e17}
\end{align}
and
\begin{align} 
D_{z} = (\tilde{D}_{1} + D_{4} + D_{p})/\sqrt{3} .\label{e17a}
\end{align}
According to the above predictions, we may conclude that the Hilbert space of the system can be split into three independent subspaces, one subspace composed of four trapping states
\begin{align}
\ket{a1} &= \frac{1}{\sqrt{2}}\left(\ket{g_{A},e_{B}} -\ket{e_{A},g_{B}}\right)\otimes\ket{1_{a},0_{b}} ,\nonumber\\
\ket{a2} &= \frac{1}{\sqrt{2}}\left(\ket{e_{A},g_{B}} -\ket{g_{A},e_{B}}\right)\otimes\ket{0_{a},1_{b}} ,\nonumber\\
\ket m &= \frac{1}{\sqrt{6}}\left\{\sqrt{2}\left(\ket{2_{a},0_{b}} + \ket{0_{a},2_{b}}\right)\otimes\ket{g_{A},g_{B}}\right. \nonumber\\
&\left. - \ket{e_{A},e_{B},0_{a},0_{b}} - \ket{g_{A},g_{B},1_{a},1_{b}}\right\} ,\nonumber\\
\ket n &=  \frac{1}{\sqrt{2}}\left(\ket{e_{A},e_{B},0_{a},0_{b}} -\ket{g_{A},g_{B},1_{a},1_{b}}\right) ,\label{eq50t}
\end{align}
and two subspaces composed of doublets
\begin{align}
\ket{w} &= \frac{1}{2}\left(\ket{g_{A},e_{B}} +\ket{e_{A},g_{B}}\right)\otimes\left(\ket{1_{a},0_{b}} - \ket{0_{a},1_{b}}\right) ,\nonumber\\
\ket{q} &= \frac{1}{\sqrt{2}}\left(\ket{2_{a},0_{b}} - \ket{0_{a},2_{b}}\right)\otimes\ket{g_{A},g_{B}} ,\label{eq51a}
\end{align}
and 
\begin{align}
\ket{u} &= \frac{1}{2}\left(\ket{g_{A},e_{B}} +\ket{e_{A},g_{B}}\right)\otimes\left(\ket{1_{a},0_{b}} + \ket{0_{a},1_{b}}\right) ,\nonumber\\
\ket{z} &= \frac{1}{\sqrt{3}}\left[\ket{e_{A},e_{B},0_{a},0_{b}} +\ket{g_{A},g_{B},1_{a},1_{b}}\right. \nonumber\\
&\left. + \frac{1}{\sqrt{2}}\left(\ket{2_{a},0_{b}} + \ket{0_{a},2_{b}}\right)\otimes\ket{g_{A},g_{B}}\right] .\label{eq51}
\end{align} 

The evolution of the doublets (\ref{eq51a}) and (\ref{eq51}) is determined by the amplitudes $(D_{w}, D_{q})$ and $(D_{u}, D_{z})$, respectively. Their time evolutions are readily found to be
\begin{align}
D_{w}(t) &= D_{w}(0)\cos\left(2gt\right) - D_{q}(0)\sin\left(2gt\right)  ,\nonumber\\
D_{q}(t) &= D_{q}(0)\cos\left(2gt\right) - D_{w}(0)\sin\left(2gt\right)  ,\label{e15n}
\end{align}
and
\begin{align}
D_{u}(t) &= D_{u}(0)\cos\!\left(2\sqrt{3}gt\right) -iD_{z}(0)\sin\!\left(2\sqrt{3}gt\right)  ,\nonumber\\
D_{z}(t) &= D_{z}(0)\cos\!\left(2\sqrt{3}gt\right) -iD_{u}(0)\sin\!\left(2\sqrt{3}gt\right)  .\label{e15}
\end{align}
Thus, due to the presence of four trapping states, the evolution of the system confines to that occurring only among states $\ket w \leftrightarrow \ket q$ and $\ket u \leftrightarrow \ket z$. Each pair of states evolves in time independently of the other, and the amplitudes of states $\ket w$ and $\ket q$ oscillate at frequency~$2g$, whereas the amplitudes of states $\ket u$ and~$\ket z$ oscillate at frequency~$2\sqrt{3}g$. 

Several features of the dynamically evolving states (\ref{eq51a}) and (\ref{eq51}) are worth noting. One can see that state $\ket w$ is a product state of a single-excitation entangled state of the atoms and the antisymmetric $N=1$ NOON state of the cavity modes~\cite{nh12,bcs89,bk00,aa10,wm11}. State $\ket q$ is a product state of the atomic state in which both atoms are in their ground states and the cavity modes being in the antisymmetric $N=2$ NOON state. This shows that an initial state in which atoms are prepared in the maximally entangled state and the cavity modes are prepared in the antisymmetric $N=1$ NOON state can be converted to the antisymmetric $N=2$ NOON state. In other words, in this scheme we can achieve a fully dynamical generation of a purely photonic $N=2$ NOON state.

The other pair of states, $\ket u$ and $\ket z$, involves the symmetric~$N=1$ and $N=2$ NOON states. One can notice that in contrast to the antisymmetric $N=1$ NOON state, an initial symmetric $N=1$ noon state cannot be completely converted to the symmetric $N=2$ NOON state. It is rather converted into the four-qubit state~$\ket z$, which involves a superposition of the~$N=2$ NOON state and two other states with the excitation evenly redistributed between the atoms or between the cavity modes. 

The reason for the difference in the transfer properties is in the trapping effect of the single excitation states. According to Eqs.~(\ref{eq14a}) and (\ref{e8p}), the antisymmetric $N=1$ NOON state is a trapping state whereas the symmetric state evolves during the atomic transit time through the cavities. Thus, the simultaneous evolution of the atomic and the symmetric NOON states results in a redistribution of the excitation between the atoms or between the cavities. 

Another interesting feature of the states (\ref{eq51a}) and (\ref{eq51}) is that the complete transfer requires the presence of an initial entanglement of the cavity modes. In other words, the complete transfer of an initial state arises from the quantum nature of the field. It is particularly seen in the form of the reduced density operator~$\rho_{AB}(t)$, which written in the computational basis is not diagonal
\begin{align}
    \rho_{AB}(t)=\left(
\begin{array}{cccc}
  \rho_{11}(t) & 0 & 0 & 0 \\
  0 & \rho_{22}(t) & \rho_{23}(t) & 0 \\
  0 & \rho_{32}(t) & \rho_{33}(t) & 0 \\
  0 & 0 & 0 & \rho_{44}(t) \\
\end{array}
\right) ,\label{11a}
\end{align}
where $\rho_{11}(t)=|D_{1}(t)|^{2}, \rho_{22}(t) =|D_{2}(t)|^{2}+|D_{5}(t)|^{2}$, $\rho_{33}(t) = |D_{3}(t)|^{2}+|D_{7}(t)|^{2}$, $\rho_{44}(t)= |D_{4}(t)|^{2}+|D_{6}(t)|^{2}+|D_{8}(t)|^{2}$, and $\rho_{23}(t) = D^{\ast}_{3}(t)D_{5}(t)+ D_{2}(t)D^{\ast}_{7}(t)$. 

The matrix (\ref{11a}) is of an $X$-state form due to the presence of the coherence $\rho_{23}(t)$ between the atoms. If the initial state of the system is taken to be one of the states~(\ref{eq51a}) or~(\ref{eq51}), then none of the trapping states (\ref{eq50t}) is populated. As a result, the coherence~$\rho_{23}(t)$ takes the form 
\begin{align}
\rho_{23}(t) = \frac{1}{2}\left(|D_{u}(t)|^{2} + |D_{w}(t)|^{2}\right) .
\end{align}
We see that the coherence is completely determined by the probabilities (populations) of states~$\ket u$ and~$\ket w$. Since the states involve maximally entangled single-quantum states of the atoms and the field modes, this implies that the coherence can be different from zero only if either of the atoms or the field modes are in an entangled state.  

The coherence is necessary for entanglement between the atoms. The concurrence evaluated from the matrix~(\ref{11a}) is given by 
\begin{equation}
    C_{AB}(t) = {\rm max}\{0,{\cal C}_{1}(t)\} ,
\end{equation}
where
\begin{align}
    {\cal C}_{1}(t) &= 2|\rho_{23}(t)| -2\sqrt{\rho_{11}(t)\rho_{44}(t)}  .
\end{align}
Clearly, ${\cal C}_{1}(t)$ can be positive indicating an entanglement between the atoms. Thus, an entanglement can be generated between the atoms without the need for introducing the auxiliary state $\ket 0$.  

We now present some numerical calculations of entanglement evolution between the atoms and the cavity modes to check if the complete transfer of the double excitation states is accompanied by the complete transfer of entanglement. We evaluate logarithmic negativities $N_{AB}(t)$ and $N_{ab}(t)$ for three different initial states. The reason to evaluate the logarithmic negativity rather than the concurrence is in the dimension of the Hilbert space of the system with two excitations present. A detailed description of the method of evaluating the logarithmic negativities is given in the Appendix. 
\begin{figure}[ht]
\centerline{\includegraphics [clip,width=3in,angle=0]{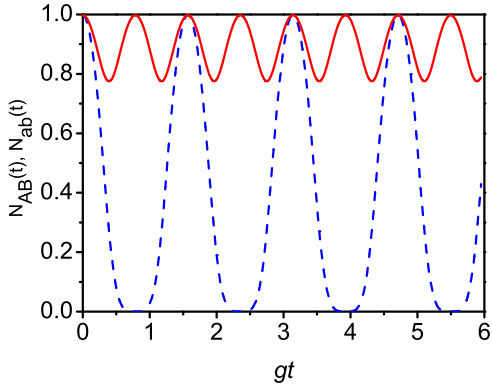}}
\caption{(Color online) Time evolution of the logarithmic negativities $N_{AB}(t)$ (blue dashed line) and $N_{ab}(t)$ (red solid line) for the initial state $\ket w$, Eq.~(\ref{eq51a}). The other parameters are $\Delta=0$ and~$g_{b}=g_{a}=g$.} 
\label{sbfig3}
\end{figure}

Figure~\ref{sbfig3} shows the logarithmic negativities $N_{AB}(t)$ and~$N_{ab}(t)$ as a function of the dimensionless time $gt$ for the initial state $\ket w$. According to Eq.~(\ref{e15n}), at times $t=n\pi/(4g),\, n=1,3,\ldots$, when state $\ket w$ is completely transferred to state $\ket q$, the atoms are expected to be separable and the field modes should be found in the maximally entangled $N=2$ NOON state. We see that at the initial time $t=0$, both negativities are maximal, $N_{AB}(0)=N_{ab}(0)=1$, indicating that initially the atoms and the field modes were maximally entangled. As time progresses, the atoms become periodically disentangled but the field modes never completely disentangle. At times $t=n\pi/(4g)$, when the atoms are separable, the modes again become maximally entangled. However, the maximal entanglement of the modes does not mean that at these times the modes returned to their initial state. At times $t=n\pi/(4g)$, the system is in the $\ket q$ state, in which the atoms are in their ground states and the field modes are in the antisymmetric~$N=2$ NOON state. 

It is interesting to note that although the evolution of the states is unitary the entanglement between the field modes oscillates twice as fast as the entanglement between the atoms. The reason is that the reduced state of the cavity modes $\rho_{ab}(t)$ is composed of two maximally entangled states, $(\ket{1_{a},0_{b}} - \ket{0_{a},1_{b}})/\sqrt{2}$ and $(\ket{2_{a},0_{b}} - \ket{0_{a},2_{b}})/\sqrt{2}$, whereas the reduced state of the atoms $\rho_{AB}(t)$ is composed of the maximally entangled state $(\ket{g_{A},e_{B}} +\ket{e_{A},g_{B}})/\sqrt{2}$ and the separable state $\ket{g_{A}g_{B}}$. Since the field and atomic states oscillate with the same Rabi frequency, $2g$, the cavity modes become maximally entangled whenever the modes are in either $(\ket{1_{a},0_{b}} - \ket{0_{a},1_{b}})/\sqrt{2}$ or $(\ket{2_{a},0_{b}} - \ket{0_{a},2_{b}})$. On the other hand, the atoms are maximally entangled only if the atoms are in one of the two atomic states, $(\ket{g_{A},e_{B}} +\ket{e_{A},g_{B}})/\sqrt{2}$.  
\begin{figure}[ht]
\centerline{\includegraphics [clip,width=3in,angle=0]{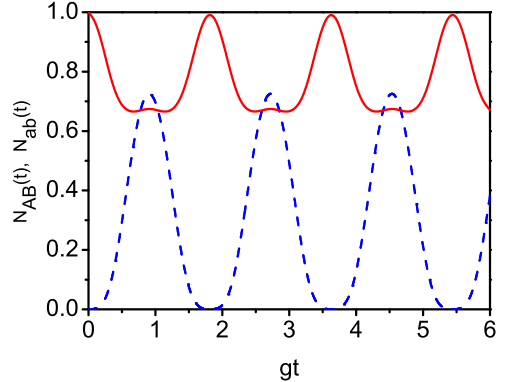}}
\caption{(Color online) Time evolution of the logarithmic negativities $N_{AB}(t)$ (blue dashed line) and $N_{ab}(t)$ (red solid line) when the atoms in their ground states $\ket{g_{A}g_{B}}$ pass through the cavities prepared initially in the maximally entangled state $(\ket{1_{a}1_{b}}+\ket{0_{a}0_{b}})/\sqrt{2}$. The other parameters are $\Delta=0$ and~$g_{b}=g_{a}=g$.} 
\label{sbfig4}
\end{figure}

If the initial state is not one of the states (\ref{eq51a}) or~(\ref{eq51}), only a partial transfer of the initial entanglement may occur. To illustrate this, we show in Fig.~\ref{sbfig4} the time evolution of the logarithmic negativities when the system starts from an initial state
\begin{eqnarray}
\ket{\Psi_{0}} = \frac{1}{\sqrt{2}}(\ket{1_{a}1_{b}}+\ket{0_{a}0_{b}})\otimes \ket{g_{A}g_{B}} ,\label{eq67}
\end{eqnarray}
in which the cavity modes are in the maximally entangled two-photon state and the atoms are in their ground states. We see that in this case the concurrence $N_{ab}(t)$ never becomes zero as time develops and $N_{AB}(t)$ never reaches the maximal value of $N_{AB}(t)=1$. Thus, no complete transfer of the entanglement occurs. The modes remain highly entangled at times when the atomic entanglement reaches its maximum value. This result is as expected from the above simple discussion that for states different than the states (\ref{eq51a}) and (\ref{eq51}), a part of the initial excitation is trapped in one of the trapping state preventing the complete transfer of the initial state to occur. It is easy to see. With the initial state (\ref{eq67}), the probabilities of the trapping states (\ref{eq50t}) are
\begin{align}
&|D_{a1}(0)|^{2} = |D_{a2}(0)|^{2} = 0 ,\nonumber\\
&|D_{m}(0)|^{2} = \frac{1}{12} ,\quad |D_{n}(0)|^{2} = \frac{1}{4} ,
\end{align}
from which it is clear that one third of the initial population is in the trapping states $\ket m$ and $\ket n$.

It is interesting to contrast the initial state (\ref{eq67}), which involves the auxiliary zero-excitation state, with an initial state involving states containing two quanta of excitation. The former was treated in the case of independent JC cells~\cite{MJ06,MJ07,MJ10,SMZ09} and shown crucial for the creation of en entanglement between the atoms.
\begin{figure}[ht]
\centerline{\includegraphics [clip,width=3.5in,angle=0]{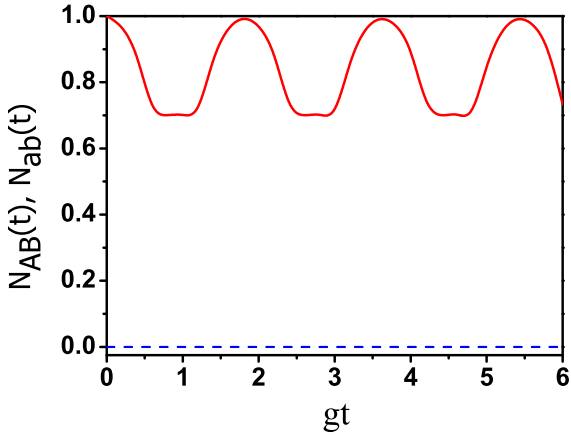}}
\caption{(Color online) Time evolution of the logarithmic negativities $N_{AB}(t)$ (blue dashed line) and $N_{ab}(t)$ (red solid line) when two atoms in their ground states pass through the cavities prepared initially in the symmetric $N=2$ NOON state: $\ket{\Psi_{0}} = (1/\sqrt{2})(\ket{2_{a}0_{b}}+\ket{0_{a}2_{b}})\otimes \ket{g_{A}g_{B}}$. The other parameters are $\Delta=0$,~$g_{b}=g_{a}=g$.} 
\label{sbfig5}
\end{figure} 

Figure~\ref{sbfig5} shows the time evolution of the logarithmic negativities for an initial state in which the atoms are in their ground states and the field modes are in the $N=2$ NOON state $(\ket{2_{a}0_{b}}+\ket{0_{a}2_{b}})/\sqrt{2}$. While Fig.~\ref{sbfig4} shows that an initial entanglement of the field modes can be transferred to the atoms when the modes are prepared in the state (\ref{eq67}), Fig.~\ref{sbfig5} shows no such transfer occurs when the modes are initially prepared in the symmetric $N=2$ NOON state.

\subsection{Transfer mediated by photons}

We now turn into the other configuration in which the coupling
between the cavities is mediated by the overlapped cavity modes.
Such coupling configuration, illustrated in Fig.~\ref{sbfig1}(b),
is represented by the interaction Hamiltonian (\ref{e3a}). What we
are going to calculate is essentially the same as for the first
scenario. We use the same basis states, Eqs.~(\ref{e7}), to find
the equations of motion for the probability amplitudes required to
study the dynamics of entanglement. As in the previous case, the
starting point is the Schr\"{o}dinger equation (\ref{e4}), which
with the Hamiltonian~(\ref{e3a}) leads to the following set of
equations of motion for the probability amplitudes
\begin{align}
i\dot{D}_{1} &= \Delta D_1+g_{b}D_2+g_{a}D_3 ,\nonumber\\
i\dot{D}_{2} &= g_{b}D_1+g_{a}D_4+\kappa D_5 ,\nonumber\\
i\dot{D}_{3} &= g_{a}D_1+g_{b}D_4+ \kappa D_7 ,\nonumber\\
i\dot{D}_{4} &= -\Delta D_4+g_{a}D_2+g_{b}D_3+\sqrt{2}\kappa (D_6+D_8)  ,\nonumber\\
i\dot{D}_{5} &=  \sqrt{2}g_{a}D_6+\kappa D_2 ,\nonumber\\
i\dot{D}_{6} &= -\Delta D_6+\sqrt{2}\kappa D_4+\sqrt{2}g_{a}D_5 ,\nonumber\\
i\dot{D}_{7} &= \kappa D_3+\sqrt{2}g_{b}D_8 ,\nonumber\\
i\dot{D}_{8} &= -\Delta D_8+\sqrt{2}\kappa D_4+\sqrt{2}g_{b}D_7 .\label{e34}
\end{align}

Before proceeding with the solution of Eqs.~(\ref{e34}), we first rearrange them to see if there are any trapping states in the system. For such a treatment it is convenient to introduce linear combinations
\begin{align} 
\tilde{D}_{2} &= (D_{2} + D_{3})/\sqrt{2} ,\quad \tilde{D}_{3} = (D_{2} - D_{3})/\sqrt{2} ,\nonumber\\
\tilde{D}_{5} &= (D_{5} + D_{7})/\sqrt{2} ,\quad \tilde{D}_{7} = (D_{5} - D_{7})/\sqrt{2} ,\nonumber\\
\tilde{D}_{6} &= (D_{6} + D_{8})/\sqrt{2} ,\quad \tilde{D}_{8} = (D_{6} - D_{8})/\sqrt{2} ,\label{e35}
\end{align}
for which Eqs.~(\ref{e34}), with $g_{a}=g_{b}\equiv g$, form two separate sets, one involving five coupled equations  
\begin{align}
i\dot{D}_{1} &= \Delta D_1+\sqrt{2}g\tilde{D}_{2} ,\nonumber\\
i\dot{\tilde{D}}_{2} &= \sqrt{2}gD_{1} +\sqrt{2}gD_{4} + \kappa \tilde{D}_{5} ,\nonumber\\
i\dot{D}_{4} &= -\Delta D_{4} +\sqrt{2}g\tilde{D}_{2} + 2\kappa\tilde{D}_{6} ,\nonumber\\
i\dot{\tilde{D}}_{5} &= \sqrt{2}g\tilde{D}_{6} + \kappa \tilde{D}_{2} ,\nonumber\\
i\dot{\tilde{D}}_{6} &= -\Delta \tilde{D}_{6} + 2\kappa D_{4} + \sqrt{2}g\tilde{D}_{5} ,\label{e36}
\end{align}
and the other involving three coupled equations
\begin{align}
i\dot{\tilde{D}}_{3} &= \kappa \tilde{D}_{7} ,\nonumber\\
i\dot{\tilde{D}}_{7} &= \kappa \tilde{D}_{3} +\sqrt{2}g\tilde{D}_{8} ,\nonumber\\
i\dot{\tilde{D}}_{8} &= -\Delta \tilde{D}_{8} +\sqrt{2}g\tilde{D}_{7} .\label{e37}
\end{align}
It is easily verified that the determinants of the coefficients of the coupled differential equations (\ref{e36}) and (\ref{e37}) are different from zero only if $\Delta\neq 0$. Thus, in the case of $\Delta\neq 0$, there are no trapping states in the system.

\subsubsection{The case of $\Delta = 0$}

Let us first specialise Eqs.~(\ref{e36}) and (\ref{e37}) to the case of exact resonance, $\Delta = 0$, in which trapping states occur. As before, the occurrence of trapping states is identified by the presence of constants of motion. It is easily verified from Eqs.~(\ref{e36}) that a linear superposition  
\begin{align}
D_{a} = \frac{1}{\Omega_{a}^{2}}\left[g^{2}D_{4} - \sqrt{2}g\kappa\tilde{D}_{5} +(\kappa^{2}-g^{2})D_{1}\right] ,
\end{align}
in which $\Omega_{a}^{2} =\sqrt{2g^{4} +\kappa^{4}}$, is a constant of motion, i.e.~$i\dot{D}_{a}=0$. 

Similarly, the system of coupled equations~(\ref{e37}) can be rewritten in terms of linear superpositions 
\begin{align}
D_{\alpha} = \frac{\sqrt{2}g}{\Omega^{\prime}}\tilde{D}_{3} - \frac{\kappa}{\Omega^{\prime}}\tilde{D}_{8} ,\quad 
D_{\beta} = \frac{\kappa}{\Omega^{\prime}}\tilde{D}_{3}  + \frac{\sqrt{2}g}{\Omega^{\prime}}\tilde{D}_{8} ,
\end{align}
where $\Omega^{\prime} =\sqrt{2g^{2} + \kappa^{2}}$, for which we find that  
\begin{align}
i\dot{D}_{\alpha} &= 0 ,\nonumber\\
i\dot{\tilde{D}}_{7} &= \Omega^{\prime} D_{\beta} ,\quad
i\dot{D}_{\beta} = \Omega^{\prime} \tilde{D}_{7} .\label{e39}
\end{align}
Thus, the amplitude $D_{\alpha}$ is a constant of motion whereas the amplitudes $\tilde{D}_{7}$ and $D_{\beta}$ undergo a simple sinusoidal evolution
\begin{align}
\tilde{D}_{7}(t) &= \tilde{D}_{7}(0)\cos\Omega^{\prime} t - D_{\beta}(0)\sin\Omega^{\prime} t  ,\nonumber\\
D_{\beta}(t) &= D_{\beta}(0)\cos\Omega^{\prime} t - \tilde{D}_{7}(0)\sin\Omega^{\prime} t  .\label{e75}
\end{align}
Evidently, states corresponding to the combinations $\tilde{D}_{7}$ and~$D_{\beta}$ can be completely transferred between themselves with frequency $\Omega^{\prime}$. 

The states corresponding to the combinations $\tilde{D}_{7}$ and~$D_{\beta}$ are of the form
\begin{align}
\ket{7} &= \frac{1}{\sqrt{2}}\left(\ket{e_{A},g_{B},1_{a},0_{b}} - \ket{g_{A},e_{B},0_{a},1_{b}}\right) ,\nonumber\\
\ket{\beta} &= \frac{\kappa}{\sqrt{2}\Omega^{\prime}}\left(\ket{e_{A},g_{B},0_{a},1_{b}} - \ket{g_{A},e_{B},1_{a},0_{b}}\right) \nonumber\\
&+\frac{g}{\Omega^{\prime}}\left(\ket{2_{a},0_{b}} - \ket{0_{a},2_{b}} \right)\!\otimes\!\ket{g_{A},g_{B}} ,\label{e76}
\end{align}
from which one can easily notice that the states contain correlations between the atoms and the field modes. 
This implies that the complete transfer of the states will not lead to the complete transfer of entanglement.

\subsubsection{The case of $\Delta\neq 0$}

Let us now examine the situation when~$\Delta\neq 0$ and focus our attention on the case of $g_{a}=g_{b}\equiv g$. With nonzero detuning the dynamics of the system becomes more complicated and simple analytical results for the probability amplitudes are not possible. However, in the limit of the strong coupling between the cavities, $\kappa\gg g$, one can uncover some interesting results. By introducing symmetric and antisymmetric  combinations 
\begin{align} 
D_{\eta} &= (D_{4} + \tilde{D}_{6})/\sqrt{2} ,\quad D_{\epsilon} = (D_{4} - \tilde{D}_{6})/\sqrt{2} ,\nonumber\\
D_{\lambda} &= (\tilde{D}_{2} + \tilde{D}_{5})/\sqrt{2} ,\quad D_{\vartheta} = (\tilde{D}_{2} - \tilde{D}_{5})/\sqrt{2} ,\label{e77}
\end{align}
and going into a rotating frame through the relations
\begin{align}
\tilde{D}_{\eta} &= D_{\eta}{\rm e}^{-i(\Delta -2\kappa)t} ,\quad \tilde{D}_{\lambda} = D_{\lambda}{\rm e}^{-i(\Delta -2\kappa)t} ,\nonumber\\
\tilde{D}_{\epsilon} &= D_{\epsilon}{\rm e}^{-i(\Delta +2\kappa)t} ,\quad \tilde{D}_{\vartheta} = D_{\vartheta}{\rm e}^{-i(\Delta +2\kappa)t} ,\nonumber\\
\tilde{D}_{1} &= D_{1}{\rm e}^{-i\Delta t} ,\label{e78}
\end{align}
we find that Eqs.~(\ref{e36}) can be written in terms of these combinations as 
\begin{align}
i\dot{\tilde{D}}_{\eta} &= \sqrt{2}g\tilde{D}_{\lambda} ,\nonumber\\
i\dot{\tilde{D}}_{\lambda} &= (\Delta -\kappa)\tilde{D}_{\lambda} +\sqrt{2}g\tilde{D}_{\eta} +\sqrt{2}g\tilde{D}_{1}{\rm e}^{2i\kappa t} ,\nonumber\\
i\dot{\tilde{D}}_{\epsilon} &= \sqrt{2}g\tilde{D}_{\vartheta} ,\nonumber\\
i\dot{\tilde{D}}_{\vartheta} &= (\Delta +\kappa)\tilde{D}_{\vartheta} +\sqrt{2}g\tilde{D}_{\epsilon} +\sqrt{2}g\tilde{D}_{1}{\rm e}^{-2i\kappa t} ,\nonumber\\
i\dot{\tilde{D}}_{1} &= g\left(D_{\lambda}{\rm e}^{2i\kappa t} +D_{\vartheta}{\rm e}^{-2i\kappa t}\right) .\label{e79}
\end{align}
We see that the terms proportional to $\tilde{D}_{1}$ play the role of a coupling between the symmetric and antisymmetric combinations. These are accompanied by terms $\exp(\pm 2i\kappa t)$ that oscillate in time with a frequency $2\kappa$. 
In the limit of $\kappa\gg g$, we can make a secular approximation, in which we ignore the rapidly oscillating terms $\exp(\pm 2i\kappa t)$. This has the effect of decoupling the equations of motion for the symmetric combinations
\begin{align}
i\dot{\tilde{D}}_{\eta} &= \sqrt{2}g\tilde{D}_{\lambda} ,\nonumber\\
i\dot{\tilde{D}}_{\lambda} &= (\Delta -\kappa)\tilde{D}_{\lambda} +\sqrt{2}g\tilde{D}_{\eta} ,\label{e80}
\end{align}
from the equations of motion for the antisymmetric combinations
\begin{align}
i\dot{\tilde{D}}_{\epsilon} &= \sqrt{2}g\tilde{D}_{\vartheta} ,\nonumber\\
i\dot{\tilde{D}}_{\vartheta} &= (\Delta +\kappa)\tilde{D}_{\vartheta} +\sqrt{2}g\tilde{D}_{\epsilon} ,\label{e81}
\end{align}
that the amplitudes of the symmetric and antisymmetric combinations evolve independently. 

It is seen that the symmetric combinations oscillate with unequal frequencies that differ by $\Delta -\kappa$, whereas the antisymmetric combinations oscillate with frequencies that differ by $\Delta +\kappa$. The difference of the frequencies will lead to oscillations of the amplitudes that are not periodic. This imperfection would result in an incomplete transfer of the states. The transfer could be complete only if the amplitudes oscillate with equal frequency. It is seen from Eq.~(\ref{e80}) that for $\Delta =\kappa$, the symmetric combinations oscillate with equal frequencies which results in sinusoidal oscillations of the amplitudes
\begin{align}
\tilde{D}_{\eta}(t) &= \tilde{D}_{\eta}(0)\cos(\sqrt{2}gt) - \tilde{D}_{\lambda}(0)\sin(\sqrt{2}gt)  ,\nonumber\\
\tilde{D}_{\lambda}(t) &= \tilde{D}_{\lambda}(0)\cos(\sqrt{2}gt) - \tilde{D}_{\eta}(0)\sin(\sqrt{2}gt)  .\label{e82}
\end{align}
Similarly, when $\Delta =-\kappa$ the antisymmetric combinations oscillate with equal frequencies and then their time evolutions~are perfectly sinusoidal
\begin{align}
\tilde{D}_{\epsilon}(t) &= \tilde{D}_{\epsilon}(0)\cos(\sqrt{2}gt) - \tilde{D}_{\vartheta}(0)\sin(\sqrt{2}gt)  ,\nonumber\\
\tilde{D}_{\vartheta}(t) &= \tilde{D}_{\vartheta}(0)\cos(\sqrt{2}gt) - \tilde{D}_{\epsilon}(0)\sin(\sqrt{2}gt)  .\label{e83}
\end{align}
Clearly, states corresponding to the symmetric combinations and those corresponding to the antisymmetric combinations of the probability amplitudes can be completely transferred between themselves. 

\begin{figure}[h]
\centerline{\includegraphics[clip,width=3in,angle=0]{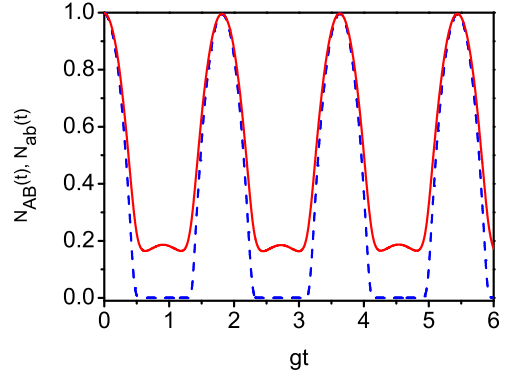}}
\caption{(Color online) Time evolution of the logarithmic negativities $N_{AB}(t)$ (blue dashed line) 
and $N_{ab}(t)$ (red solid line) for the case in which the system was initially prepared in the symmetric state $\ket\lambda$. The other parameters are $g_{b}=g_{a}=g$ and $\Delta=\kappa=10g$.}
\label{sbfig6}
\end{figure}

The states corresponding to the symmetric combinations $\tilde{D}_{\eta}$ and~$\tilde{D}_{\lambda}$ can be written in  the form
\begin{align}
\ket{\eta} &= \frac{1}{2}\left(\sqrt{2}\ket{1_{a},1_{b}} + \ket{2_{a},0_{b}} + \ket{0_{a},2_{b}} \right)\!\otimes\!\ket{g_{A},g_{B}} ,\nonumber\\
\ket{\lambda} &= \frac{1}{2}\left(\ket{g_{A},e_{B}} +\ket{e_{A},g_{B}}\right)\!\otimes\!\left(\ket{1_{a},0_{b}} + \ket{0_{a},1_{b}}\right) ,\label{e84}
\end{align}
whereas the states corresponding to the antisymmetric combinations~$\tilde{D}_{\epsilon}$ and~$\tilde{D}_{\vartheta}$ can be written as follows
\begin{align}
\ket{\epsilon} &= \frac{1}{2}\!\left(\sqrt{2}\ket{1_{a},1_{b}} - \ket{2_{a},0_{b}} - \ket{0_{a},2_{b}} \right)\!\otimes\!\ket{g_{A},g_{B}}  ,\nonumber\\
\ket{\vartheta} &= \frac{1}{2}\!\left(\ket{g_{A},e_{B}} - \ket{e_{A},g_{B}}\right)\!\otimes\!\left(\ket{1_{a},0_{b}} - \ket{0_{a},1_{b}}\right) .\label{e85}
\end{align} 
Notice an important property of states (\ref{e84}) and (\ref{e85}) that they do not contain correlations between the atoms and the field modes. The states are of the form of product states of the atomic states and those of the field states. This is in distinct contrast to the states for $\Delta =0$, which contain correlations between the atoms and the field modes. An another interesting property of states (\ref{e84}) and (\ref{e85}) is that the atoms may completely disentangle during the evolution of the states, in contrast to the field modes that can never disentangle. This is illustrated in Fig.~\ref{sbfig6} which shows the time evolution of the logarithmic negativities when the system was initially prepared in state $\ket\lambda$. We see that the initial entanglement between the atoms and that between the field modes recurs periodically. The atomic entanglement periodically falls to zero in contrast to the entanglement between the cavity mode that never ceases. 

In closing this section, we would like to point out that the two coupling configurations between the cavities lead to quite different results for transferring double excitation states. Comparing Eqs.~(\ref{eq51a}) and (\ref{eq51}), the quantum states that are completely transferred when the transfer is mediated by the atoms with Eqs.~(\ref{e84}) and (\ref{e85}) for the states that are maximally transferred when the coupling is mediated by photons, we see that the coupling mediated by the atoms may result in a "nonclassical" $N=2$ NOON state~$\ket q$, whereas no such state is created by the coupling mediated by photons. In the latter, two states are created,~$\ket\eta$ and $\ket\epsilon$ that resemble very much a two-photon "classical" state created at an asymmetric beam-splitter with unequal reflection and transmission coefficients~\cite{mw95}.

\section{Effect of losses}~\label{loss}

We now briefly discuss the effect of losses, spontaneous emission of the atoms and the cavity damping on the transfer of quantum states and entanglement. To simplify the calculations we consider only the case in which both atoms are damped with the same spontaneous emission rate $\Gamma$ and the cavity modes are damped with the same cavity leakage rate $\gamma$.
If the effect of both of these types of losses is taken into account, the free Hamiltonian $\hat{H}_{0}$ for the two configurations can be replaced by an effective Hamiltonian
\begin{align}
\hat{H}_{{\rm eff}} = \hat{H}_{0} -
\frac{i\hbar}{2}\left[\Gamma\left(\hat{\sigma}_{A}^{+}\hat{\sigma}_{A}^{-}+\hat{\sigma}_{B}^{+}\hat{\sigma}_{B}^{-}\right) + \gamma\left(\hat{a}^{\dag}\hat{a}+\hat{b}^{\dag}\hat{b}\right)\right] .\label{e86} 
\end{align}
After taking into account the effect of losses we find that the single excitation case described in Eq.~(\ref{e8p}) can be written in the form
\begin{align}
\dot{\tilde{W}}(t) &=  -\sqrt{2}ig_{0}\tilde{X}(t){\rm e}^{-\frac{1}{2}(\gamma -\Gamma)t} ,\nonumber\\
\dot{\tilde{X}}(t) &= i\Delta \tilde{X}(t) -\sqrt{2}ig_{0}\tilde{W}(t){\rm e}^{\frac{1}{2}(\gamma -\Gamma)t} ,\nonumber \\
\dot{\tilde{Y}}(t) &= i\Delta \tilde{Y}(t) ,\quad \dot{\tilde{U}}(t) =  0 ,\label{e87}
\end{align}
where $\tilde{W}=W\exp(\frac{1}{2}\Gamma t)$, $\tilde{Y}=Y\exp(\frac{1}{2}\Gamma t)$, $\tilde{U}=U\exp(\frac{1}{2}\gamma t)$, and $\tilde{X}=X\exp(\frac{1}{2}\gamma t)$. As we can see from these expressions, the two state evolution is preserved that an initial excitation can be transferred between states $\ket w$ and $\ket x$ only. The effect of the losses is to modify the Rabi frequency of the oscillations through the time dependent oscillatory terms $\exp[\pm\frac{1}{2}(\gamma-\Gamma)t]$ and to degrade the probability amplitudes of the states. In the case of $\gamma \approx \Gamma$ the losses do not change the Rabi frequency of the oscillations. Consequently, an initial entanglement is transferred between the two states only but with the magnitude degrading in time with the rate $\Gamma$. A similar result is obtained in the case of double excitation. 
\begin{figure}[ht]
\includegraphics[width=0.7\columnwidth]{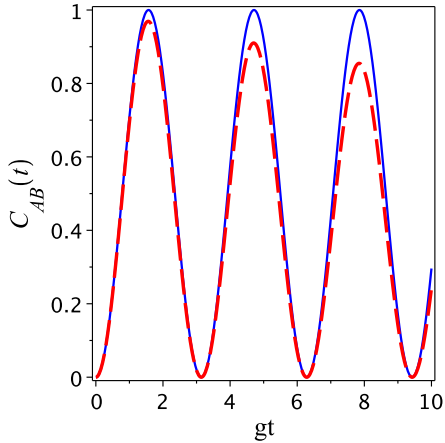}
\caption{(Color online) The time evolution of the concurrence $C_{AB}(t)$ in the presence (solid line) and the absence (dashed line) of the losses for $g_{a}=g_{b}=g$, $\Delta=\kappa$, and $\Gamma=\gamma=0.02g$. The system was initially in the state (\ref{eq12s}).}
\label{sbfig7}
\end{figure}

An example for the effect of the losses on the time evolution of the concurrence $C_{AB}(t)$ for the single excitation case is shown in Fig.~\ref{sbfig7}. The chosen parameters value of $\Gamma =\gamma =0.02g$ is consistent with current experiments involving a high-$Q$ cavity~\cite{mm05,th05} with cavity damping rate given by $\gamma =2.3\times 10^{-2}g$. It is seen from the figure that in the presence of the losses the periodic maxima of the concurrence are reduced in magnitude at the rate $\Gamma$ as time progresses. The frequency of the oscillations is not affected by the losses and non-zero concurrence persists over many Rabi periods. 
\begin{figure}[ht]
\includegraphics[width=0.7\columnwidth]{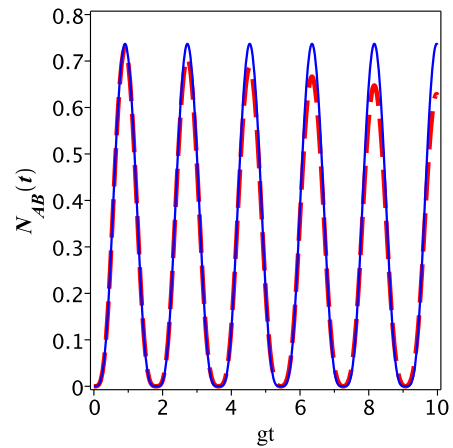}
\caption{(Color online) Time evolution of the logarithmic negativities $N_{AB}(t)$ in the presence (solid line) and the absence (dashed line) of the losses when the atoms in their ground states $\ket{g_{A}g_{B}}$ pass through the cavities prepared initially in the maximally entangled state $(\ket{1_{a}1_{b}}+\ket{0_{a}0_{b}})/\sqrt{2}$. The other parameters are $\Delta=0$, $g_{b}=g_{a}=g$, and $\Gamma =\gamma=0.02g$.}
\label{sbfig8}
\end{figure}

Figure~\ref{sbfig8} shows the effect of the losses on the time evolution of the negativity $N_{AB}(t)$ in the case of a double excitation. One can see from the figure that similar to the case of a single excitation, the effect of the losses is to reduce the magnitude of the entanglement. The periodic maxima of the concurrence decay exponentially in time at the rate $\Gamma$. 
Therefore, we can conclude that in both cases of the excitation, the two state evolution is preserved with the effect of the losses to degrade the magnitude of an initial entanglement.

\section{conclusions}~\label{seccon}

The problem of the complete transfer of quantum states has been examined in a four-qubit system composed of two single-mode cavities and two two-level atoms. We have considered two different coupling configurations between the qubits specified by two distinctly different types of the interaction Hamiltonians. In the first, the coupling is mediated by "flying" atoms that simultaneously couple to the field modes of the cavities, whereas in the second configuration, each atom resides inside one of the cavities and the coupling between the cavities is mediated by the overlapping field modes. 

We have demonstrated that the problem of the complete transfer of states is indeed tractable, that a proper choice of the basis states for the state vector of the system makes it possible to confine the dynamics of the system to that occurring between two states only. In other words, we have found that the time evolution of the state vector of the system can be divided into two state pairs that evolve in time plus time stationary trapping states.

Moreover, we have found that the complete transfer of a quantum state does not necessary mean perfect transfer of entanglement. An initial entanglement can be perfectly and reversibly transferred between the atoms and the field modes only if the states are in the form of product states of atomic and field states. If the states contain correlations between the atoms and the field modes, no perfect transfer of the entanglement is possible. 
 
We have also shown that apart from the similarities in the dynamics of the system for the two configurations there are some important differences. In particular, the complete transfer of quantum states is more manifested in the system where the coupling is mediated by the overlapping field modes. In this case not only the symmetric but also the antisymmetric states can be completely transferred between themselves. However, a purely photonic $N=2$ NOON state can be created only if the coupling is mediated by the atoms.


\appendix
\section{Concurrence and logarithmic negativity of the atomic and field systems}~\label{app}

In this Appendix we provide some details of the evaluation of the concurrence and the logarithmic negativity which are entanglement monotones~\cite{vi00,vw02,pl05}. The concurrence and the logarithmic negativity are evaluated from the knowledge of the reduced density matrices of the atomic and the field mode subsystems.

To calculate the concurrence measure of entanglement between the atoms we trace out the cavity field part of the density matrix of the total system to obtain the reduced density matrix $\rho_{AB}$ and then calculate~\cite{W98} 
\begin{equation}
\label{maxim}
C =\rm{max}\left\{0,\sqrt{\lambda_1}-\sqrt{\lambda_2}-\sqrt{\lambda_3}-\sqrt{\lambda_4}\right\} ,
\end{equation}
where $\lambda_i$ are the eigenvalues (in descending order) of the Hermitian matrix $R=\rho_{AB}\tilde{\rho}_{AB}$ in which $\tilde{\rho}_{AB}$ is given by
\begin{equation}
\tilde{\rho}_{AB} = \sigma_y\otimes\sigma_y\rho^{\ast}_{AB}\sigma_y\otimes\sigma_y.
\end{equation}
and $\sigma_y$ is a Pauli matrix. 

In the case of double excitation, the reduced density operator $\rho_{AB}$ is given by
\begin{align}
\rho_{AB} &= {\rm Tr}_{ab}(\rho) = \bra {0_{a}0_{b}} \rho \ket {0_{a}0_{b}} +\bra{1_{a}0_{b}}\rho\ket{1_{a}0_{b}} \nonumber\\
&+\bra {0_{a}1_{b}}\rho \ket{0_{a}1_{b}} 
+\bra{1_{a}1_{b}}\rho\ket{1_{a}1_{b}} \nonumber\\
&+\bra {2_{a}0_{b}} \rho\ket {2_{a} 0_{b}} +\bra {0_{a} 2_{b}} \rho\ket {0_{a} 2_{b}} .
\end{align}
In the basis spanned by four state vectors, $\ket{e_{A}e_{B}},\ket{e_{A}g_{B}},\ket {g_{A}e_{B}}, \ket{g_{A}g_{B}}$, the density operator~$\rho_{AB}$ is of the form
\begin{align}
    \rho_{AB} =\left(
\begin{array}{cccc}
  \rho_{11} & 0 & 0 & 0 \\
  0 & \rho_{22} & \rho_{23} & 0 \\
  0 & \rho_{32} & \rho_{33} & 0 \\
  0 & 0 & 0 & \rho_{44} \\
\end{array}
\right) ,\label{11u}
\end{align}
where $\rho_{11} =|D_{1}|^{2}, \rho_{22} =|D_{2}|^{2}+|D_{5}|^{2}$, $\rho_{33} = |D_{3}|^{2}+|D_{7}|^{2}$, $\rho_{44}= |D_{4}|^{2}+|D_{6}|^{2}+|D_{8}|^{2}$, $\rho_{23} = D^{\ast}_{3}D_{5} + D_{2}D^{\ast}_{7}$. 

The concurrence for the density matrix~(\ref{11u}) is given~by 
\begin{equation}
    C_{AB}(t) = {\rm max}\{0,{\cal C}_{1}(t)\} ,
\end{equation}
where
\begin{align}
    {\cal C}_{1}(t) &= 2|\rho_{23}(t)| -2\sqrt{\rho_{11}(t)\rho_{44}(t)} .
\end{align}
It is seen that ${\cal C}_{1}(t)$ can be positive indicating an entanglement between the atoms.

Another measure of entanglement, the logarithmic negativity is defined as
\begin{equation}
    N_{AB}(t) = {\rm max}\left\{0,\log_{2}[1-2\mu_{2}(t)]\right\} ,
\end{equation}
where
\begin{align}
\mu_{2}(t) &= \frac{1}{2}\left[\rho_{11}(t) + \rho_{44}(t)\right. \nonumber\\
&\left. -\sqrt{[\rho_{11}(t) - \rho_{44}(t)]^{2} + 4|\rho_{23}(t)|^2}\right] .
\end{align}
We see that $\mu_{2}(t)$ can be negative and negative values are possible only if $|\rho_{23}(t)|\neq 0$.


If one is interested in entanglement between the cavity modes, the reduced density matrix for the cavity modes $\rho_{ab}$ is obtained by tracing out the atomic part of the density matrix $\rho$. This gives 
\begin{align}
\rho_{ab} &= {\rm Tr}_{AB}(\rho) = \bra {e_{A}e_{B}} \rho\ket{e_{A}e_{B}} 
+\bra {e_{A}g_{B}}\rho\ket{e_{A}g_{B}} \nonumber\\
&+\bra{g_{A}e_{B}}\rho\ket{g_{A}e_{B}} + \bra {g_{A}g_{B}}\rho\ket{g_{A}g_{B}} .
\end{align}
However, the determination of entanglement between the cavity modes is a slightly more complicated than for the atoms. The reason is that in the case of two excitations present in the system, the subspace of the cavity modes is spanned by nine states, $\{\ket{0_{a}0_{b}},\ket{0_{a}1_{b}}, \ket {0_{a}2_{b}},\ket{1_{a}0_{b}}, \ket{1_{a}1_{b}},\ket{1_{a}2_{b}},\ket{2_{a}0_{b}},$ $\ket{2_{a}1_{b}},\ket{2_{a}2_{b}}\}$. In this basis, the reduced density matrix $\rho_{ab}$ has the form
\begin{align}\label{A10}
    \rho_{ab}(t)=\left(%
\begin{array}{ccccccccc}
  \rho_{11}& 0 & \rho_{13}& 0 & \rho_{14}&0&\rho_{17}&0&0 \\
  0 & \rho_{22} & 0 & \rho_{24}&0 & 0 & 0&0&0\\
 \rho_{31} & 0 & \rho_{33} & 0 & \rho_{35}&0& \rho_{36}&0&0 \\
  0& \rho_{42} & 0 &\rho_{44}&0&0&0&0&0 \\
 \rho_{51}&0&\rho_{53}&0&\rho_{55}&0&\rho_{57}&0&0\\
0&0&0&0&0&0&0&0&0\\
  \rho_{71}&0&\rho_{73}&0&\rho_{75}&0&\rho_{77}&0&0\\
  0&0&0&0&0&0&0&0&0\\
  0&0&0&0&0&0&0&0&0\\
\end{array}%
\right) .
\end{align}
A partial transpose $\rho_{ab}^{T_{B}}$ of the matrix (\ref{A10}) reads
\begin{align}\label{A11}
   \rho_{ab}^{T_{B}}(t)=\left(%
\begin{array}{ccccccccc}
  \rho_{11}& 0 & \rho_{31}& 0 & \rho_{24}&0&\rho_{17}&0&\rho_{37} \\
  0 & \rho_{22} & 0 & \rho_{15}&0 & \rho_{35} & 0&0&0\\
 \rho_{13} & 0 & \rho_{33} & 0 & 0&0& 0&0&0 \\
  0& \rho_{51} & 0 &\rho_{44}&0&0&0&\rho_{57}&0 \\
 \rho_{42}&0&0&0&\rho_{55}&0&0&0&0\\
0&\rho_{53}&0&0&0&0&0&0&0\\
  \rho_{71}&0&0&0&0&0&\rho_{77}&0&0\\
  0&0&0&\rho_{75}&0&0&0&0&0\\
  \rho_{73}&0&0&0&0&0&0&0&0\\
\end{array}%
\right) ,
\end{align}
where $\rho_{11}=|D_{1}|^{2}+|D_{0}|^{2},
\rho_{22}=|D_{2}|^{2}+|D_{7}|^{2}, \rho_{33}=|D_{8}|^{2},
\rho_{44}=|D_{3}|^{2}+|D_{5}|^{2}, \rho_{55}=|D_{4}|^{2},
\rho_{77} = |D_{6}|^{2},
\rho_{13}=\rho_{31}^{\ast}=D_{0}D^{\ast}_{8},
\rho_{15}=\rho_{51}^{\ast}=D_{0}D^{\ast}_{4},
\rho_{17}=\rho_{71}^{\ast}=D_{0}D^{\ast}_{6},
\rho_{24}=\rho_{42}^{\ast}=D_{2}D^{\ast}_{5}+D_{7}D^{\ast}_{3},
\rho_{35}=\rho_{53}^{\ast}=D_{8}D^{\ast}_{4},
\rho_{37}=\rho_{73}^{\ast}=D_{8}D^{\ast}_{6}$ and
$\rho_{57}=\rho_{75}^{\ast}=D_{4}D_{6}^{\ast}$.

The large dimension of the matrix (\ref{A10}) requires the use of the logarithmic negativity rather than the concurrence in the searching for entanglement between the modes. The logarithmic negativity $N_{ab}$ for the reduced density matrix~(\ref{A10}) reads
\begin{equation}
    N_{ab} = {\rm max}\{0,\log_{2}(1+2\sum_{i}|\mu_{i}|)\} ,
\end{equation}
where $\mu_{i}$ are the negative eigenvalues of the partial transpose matrix (\ref{A11}).

\end{document}